\documentclass[acmsmall]{acmart}
\AtBeginDocument{%
  }

\setcopyright{cc}
\setcctype{by}
\acmDOI{10.1145/3832204}
\acmYear{2026}
\acmJournal{PACMSE}
\acmVolume{3}
\acmNumber{ISSTA}
\acmArticle{ISSTA113}
\acmMonth{10}
\acmSubmissionID{issta26main-p997-p}
\received{2026-01-30}
\received[accepted]{2026-06-25}

\usepackage{algorithmic}
\usepackage{graphicx}
\usepackage{textcomp}
\usepackage{xcolor}
\usepackage{tabularx}
\usepackage{multirow}
\usepackage{booktabs}
\usepackage{array}
\usepackage{threeparttable}
\usepackage{subcaption}
\usepackage{soul}
\usepackage{url}
\usepackage{enumitem}
\usepackage{listings}
\usepackage{minted}
\usepackage{pifont} 
\usepackage[most]{tcolorbox}
\usepackage[table]{xcolor}
\usepackage{array}
\usepackage{ragged2e}

\definecolor{PromptLabel}{HTML}{E5E7EB}
\definecolor{PromptShared}{HTML}{F3F4F6}
\definecolor{PromptBase}{HTML}{EAF2FF}
\definecolor{PromptAdvanced}{HTML}{FFF3CD}
\definecolor{PromptReasoning}{HTML}{F3E8FF}
\definecolor{PromptBaseModel}{HTML}{FCE7F3}

\newcolumntype{L}{>{\bfseries\tiny\RaggedRight\arraybackslash}p{0.20\textwidth}}
\newcolumntype{Z}{>{\tiny\RaggedRight\arraybackslash}X}

\newcommand{\promptcode}[1]{\texttt{\tiny #1}}

\newcommand{\prompttag}[2]{%
    \begingroup
    \setlength{\fboxsep}{1pt}%
    \colorbox{#1}{\strut\textbf{#2}}%
    \endgroup
}

\newcommand{\promptcomponent}[2]{%
    \parbox[t]{\linewidth}{%
        \textsc{#1}\par\vspace{-0.15em}%
        {\normalfont\textcolor{black!60}{(#2)}}%
    }%
}

\newcommand{\bt}{\textasciigrave}
\newcommand{\javafence}{\bt\bt\bt{}java}
\newcommand{\mdfence}{\bt\bt\bt}

\definecolor{QualLineNo}{HTML}{9AA0A6}
\definecolor{QualDocText}{HTML}{666A70}
\definecolor{QualKeyword}{HTML}{B12AC2}
\definecolor{QualType}{HTML}{E85D75}
\definecolor{QualString}{HTML}{3A9D45}
\definecolor{QualMethod}{HTML}{2F80ED}

\newtcolorbox{qualfigurebox}{
    enhanced,
    colback=white,
    colframe=white,
    boxrule=0pt,
    arc=0pt,
    left=1pt,
    right=1pt,
    top=0pt,
    bottom=0pt,
    boxsep=0pt,
    width=\linewidth,
    before skip=0pt,
    after skip=0pt
}

\newtcolorbox{qualcodebox}[1][]{
    enhanced,
    colback=white,
    colframe=white,
    boxrule=0pt,
    arc=0pt,
    left=0pt,
    right=1pt,
    top=0.25pt,
    bottom=0.25pt,
    boxsep=0pt,
    fontupper=\ttfamily\tiny,
    #1
}

\newcommand{\qualtitle}[2][\linewidth]{%
    \tcbox[
        on line,
        enhanced,
        colback=black,
        colframe=black,
        boxrule=0pt,
        arc=0pt,
        left=2.2pt,
        right=2.2pt,
        top=0.9pt,
        bottom=0.9pt,
        boxsep=0pt
    ]{\sffamily\bfseries\tiny\color{white}\vphantom{Agjpqy}#2}%
    \par\vspace{-0.15em}%
}

\newcommand{\qualdivider}{%
    \vspace{0.25em}
    {\color{black!18}\hrule height 0.25pt}
    \vspace{0.25em}
}

\newcommand{\qline}[2]{%
    \makebox[1.05em][r]{\textcolor{QualLineNo}{#1}}\hspace{0.25em}#2\par
}

\newcommand{\qlinebadge}[3]{%
    \makebox[1.05em][r]{\textcolor{QualLineNo}{#1}}\hspace{0.25em}#2\hfill#3\par
}

\newcommand{\qkw}[1]{\textcolor{QualKeyword}{#1}}
\newcommand{\qtype}[1]{\textcolor{QualType}{#1}}
\newcommand{\qstr}[1]{\textcolor{QualString}{#1}}
\newcommand{\qmeth}[1]{\textcolor{QualMethod}{#1}}
\newcommand{\qdoc}[1]{\textcolor{QualDocText}{#1}}

\newcommand{\badgebad}[1]{%
    \tcbox[
        on line,
        colback=red!16,
        colframe=red!18,
        boxrule=0.2pt,
        arc=1pt,
        left=1pt,
        right=1pt,
        top=0.25pt,
        bottom=0.25pt,
        boxsep=0pt
    ]{\sffamily\tiny\color{red!75!black}#1 \ding{55}}%
}

\newcommand{\badgegood}[1]{%
    \tcbox[
        on line,
        colback=green!16,
        colframe=green!20,
        boxrule=0.2pt,
        arc=1pt,
        left=1pt,
        right=1pt,
        top=0.25pt,
        bottom=0.25pt,
        boxsep=0pt
    ]{\sffamily\tiny\color{green!45!black}#1 \ding{51}}%
}

\soulregister{\cite}{7}
\soulregister{\citeauthor}{7}
\soulregister{\footnote}{7}
\soulregister{\ref}{7}
\soulregister{\st}{7}
\soulregister{\subsection}{7}
\soulregister{\textbf}{7}
\soulregister{\textsc}{7}

\lstdefinestyle{promptstyle}{
    backgroundcolor=\color{white},    
    frame=single,                         
    breaklines=true,                    
    breakatwhitespace=true,             
    breakindent=10pt,                   
    basicstyle=\tiny\ttfamily,  
    showstringspaces=false,             
    keepspaces=true,                    
    numberstyle=\tiny\color{black!50},  
    stepnumber=1,
    numbersep=5pt,
    tabsize=2,
}

\usepackage[most]{tcolorbox}   

\makeatletter
\let\bvf@old@SwitchCases\lstKV@SwitchCases
\def\lstKV@SwitchCases#1#2#3{}
\usepackage{lstlinebgrd}
\let\lstKV@SwitchCases\bvf@old@SwitchCases
\makeatother

\definecolor{bvfHL}     {RGB}{255, 240, 180}  
\definecolor{bvfKwd}    {RGB}{  0,  51, 153}  
\definecolor{bvfStr}    {RGB}{139,  30,  30}  
\definecolor{bvfCom}    {RGB}{80, 92, 104}  
\definecolor{bvfFixed}  {RGB}{ 20, 120, 110}  
\definecolor{bvfFixedBg}{RGB}{ 15,  95,  90}  
\definecolor{bvfBuggy}  {RGB}{170,  40,  40}  
\definecolor{bvfBuggyBg}{RGB}{140,  30,  30}  

\newcommand{\hlRange}[2]{%
  \ifnum\value{lstnumber}<#1 \else
  \ifnum\value{lstnumber}>#2 \else
    \color{bvfHL}%
  \fi\fi
}

\newcommand{\hltok}[1]{%
  \begingroup
    \setlength{\fboxsep}{0.5pt}%
    \colorbox{bvfHL}{#1}%
  \endgroup
}

\lstdefinestyle{codeexample}{
  language=Java,
  basicstyle=\ttfamily\footnotesize,
  keywordstyle=\color{bvfKwd}\bfseries,
  commentstyle=\color{bvfCom}\itshape,
  stringstyle=\color{bvfStr},
  showstringspaces=false,
  columns=fullflexible,
  keepspaces=true,
  tabsize=4,
  breaklines=false,
  aboveskip=2pt,
  belowskip=2pt,
  xleftmargin=0pt,
  xrightmargin=0pt,
  framesep=0pt,
  frame=none,
}

\lstdefinestyle{codeTiny}{
  style=codeexample,
  basicstyle=\ttfamily\tiny,
  aboveskip=0pt,
  belowskip=0pt,
}

\tcbset{
  compactfulltitle/.style={
    enhanced,
    sharp corners,
    boxrule=0.4pt,
    colback=white,
    fonttitle=\bfseries\scriptsize,
    coltitle=white,
    left=0pt, right=0pt,
    top=0pt, bottom=0pt,
    toptitle=-0.5pt,
    bottomtitle=-1pt,
    lefttitle=0.5pt,
    righttitle=0.5pt,
  },
  fixedbox/.style={
    compactfulltitle,
    colframe=teal!60!black,
    colbacktitle=teal!70!black,
  },
  buggybox/.style={
    compactfulltitle,
    colframe=red!60!black,
    colbacktitle=red!70!black,
  },
  docbox/.style={
    compactfulltitle,
    colframe=violet!60!black,
    colbacktitle=violet!70!black,
  },
  docstringbox/.style={
    compactfulltitle,
    colframe=orange!70!black,
    colbacktitle=orange!75!black,
  },
}

\newsavebox{\falbox}

\begin{document}

\title{Evaluating and Mitigating the Misguidance Effect of Buggy Code in LLM-Generated Unit Tests}

\author{Junda Zhao}
\orcid{0000-0003-4978-4128}
\affiliation{%
  \institution{University of Toronto}
  \department{Department of Mechanical and Industrial Engineering}
  \city{Toronto}
  \country{Canada}
}
\email{junda.zhao@mail.utoronto.ca}

\author{Shurui Zhou}
\orcid{0000-0002-6346-6073}
\affiliation{%
  \institution{University of Toronto}
  \department{Department of Electrical and Computer Engineering}
  \city{Toronto}
  \country{Canada}
}
\email{shurui.zhou@utoronto.ca}

\author{Eldan Cohen}
\orcid{0000-0001-5767-6683}
\affiliation{%
  \institution{University of Toronto}
  \department{Department of Mechanical and Industrial Engineering}
  \city{Toronto}
  \country{Canada}
}
\email{eldan.cohen@utoronto.ca}

\renewcommand{\shortauthors}{Junda Zhao, Shurui Zhou, and Eldan Cohen}

\begin{abstract}
While Large Language Models (LLMs) show great promise for automating unit test generation, recent studies suggest that the quality of generated tests can be negatively impacted when models are prompted with buggy code. This paper presents a new metric to quantitatively measure the ``misguidance effect,'' a phenomenon where buggy code steers LLMs toward generating tests that validate its erroneous behavior rather than expose it. Our analysis reveals that prompting LLMs with buggy code has a severe, twofold impact: it significantly increases ``misguided tests'' that assert incorrect behavior while simultaneously suppressing the generation of effective, bug-finding tests. We further corroborate this effect from a model-internal perspective, showing that buggy code skews LLMs' preference toward tests that assert the same erroneous behavior. To counter this, we introduce and validate a specification-based unit test generation paradigm that replaces the code under test in the prompt with an LLM-generated specification docstring. Our results show that this paradigm effectively reduces misguided tests while substantially increasing effective tests, improves multi-round, feedback-driven test generation pipelines, and remains applicable to both buggy and bug-free code. Overall, these results suggest that specification-based prompting is a promising strategy for mitigating misguidance from buggy code in LLM-generated unit tests.

\end{abstract}





\maketitle

\section{Introduction}
Unit testing plays a pivotal role in software quality assurance, verifying that individual components behave as expected \cite{sommerville2011software}. By detecting defects early in the development life-cycle, unit tests help reduce maintenance costs, facilitate refactoring, and enhance overall software reliability \cite{SoftwareTestingWithLLMSurvey}. Despite these benefits, creating manual tests remains labor-intensive and error-prone, motivating efforts to automate the process \cite{SurveyOnUnitTestPracticesandProblems}.

Recent advancements in Large Language Models (LLMs) have yielded powerful models such as GPT \cite{openai2024gpt4ocard}, DeepSeek \cite{deepseekai2025deepseekv3technicalreport}, and Claude \cite{anthropic2024claude35sonnet}, which demonstrated impressive capabilities across a range of code-related tasks and prompted several recent studies focused on leveraging and evaluating their performance in automating unit test generation \cite{ChatUniTestFSE2024Demo, EvalImprChatgptUTGenFSE2024, TestGenLLMandMutIST, OnLLMUTGenEvalASE2024, EmpEvalforLLMUTAutoGenTSE}.

Existing studies on LLM-based unit test generation have predominantly used bug-free code as the primary prompt input and often relied on metrics like test correctness and coverage \cite{ChatUniTestFSE2024Demo, EvalImprChatgptUTGenFSE2024, TestGenLLMandMutIST, OnLLMUTGenEvalASE2024, EmpEvalforLLMUTAutoGenTSE}, which do not directly reflect the bug detection effectiveness of the generated tests \cite{CovNotStronglyCorrWithEffICSE2014}. This common experimental design, however, fails to mirror real-world scenarios where the code under test is often buggy, which may negatively impact the effectiveness of the generated test suite. To date, only a small number of works consider buggy code as input: Abdullin et al. \cite{abdullin2025testwarscomparativestudy} report on bug detection effectiveness when generating tests from buggy sources; Mathews et al. \cite{buggycodeprevent} argue that buggy code can impede its own detection in automatic test generation systems; and He et al. \cite{he2025usepropertybasedtestingbridge} find that, in iterative code–test generation, tests derived from buggy code generated in previous iterations may reinforce existing flaws in the code. Despite the emerging awareness, these studies neither quantify the performance degradation compared to bug-free code input nor establish the underlying source of this performance degradation. 

A recent study by Huang et al. \cite{huang2025measuringinfluenceincorrectcode} has attempted to quantitatively measure the extent to which buggy code misguides LLMs when generating unit tests—i.e., leads models to treat the erroneous behavior of buggy code as intended functionality and to generate tests that validate such behavior. However, their metric for quantifying this misguidance effect has a critical limitation: it labels all tests that pass on the buggy version as ``misguided.'' In practice, buggy code often behaves similarly to correct code on many inputs and only differs on specific inputs that trigger the bug. In fact, as we show empirically in Section~\ref{subsec:metrics}, the majority of tests generated for buggy code that successfully pass on the buggy version also pass on its fixed counterpart, indicating that the metric proposed by Huang et al. fails to capture the existence and extent of the misguidance phenomenon accurately.

\begin{figure*}[htbp!]
\vspace{-0.6em}
\centering
\begin{minipage}{1.00\textwidth}

\begin{minipage}[t]{0.333\linewidth}
\begin{tcolorbox}[fixedbox, title=Bug-free method,
                  equal height group=row1]
\begin{lstlisting}[style=codeTiny,
    linebackgroundcolor={\hlRange{6}{11}}]
public static boolean equals(
 CharSequence cs1, CharSequence cs2){
  if (cs1 == cs2)       return true;
  if (cs1 == null
      || cs2 == null)   return false;
  if (cs1 instanceof String
      && cs2 instanceof String)
    return cs1.equals(cs2);
  return CharSequenceUtils
   .regionMatches(cs1, false, 0, cs2, 0, 
    Math.max(cs1.length(), cs2.length()));
}
\end{lstlisting}
\end{tcolorbox}
\end{minipage}%
\hfill
\begin{minipage}[t]{0.310\linewidth}
\begin{tcolorbox}[buggybox, title=Buggy method,
                  equal height group=row1]
\begin{lstlisting}[style=codeTiny,
    linebackgroundcolor={\hlRange{6}{6}}]
public static boolean equals(
 CharSequence cs1, CharSequence cs2){
  if (cs1 == cs2)       return true;
  if (cs1 == null
      || cs2 == null)   return false;
  return cs1.equals(cs2);
}
\end{lstlisting}
\end{tcolorbox}
\end{minipage}%
\hfill
\begin{minipage}[t]{0.349\linewidth}
\begin{tcolorbox}[docbox, title=Docstring generated from buggy method,
                  equal height group=row1]
\begin{lstlisting}[style=codeTiny]
/**
 * Compares two CharSequence instances for
 * equality, handling null inputs safely.
 *
 * This method provides a null-safe, case-
 * sensitive comparison. Two CharSequence
 * objects are considered equal if they re-
 * present the same sequence of characters.
 * The method is designed to work with any
 * CharSequence implementation, such as
 * String, StringBuilder, or StringBuffer.
 */
\end{lstlisting}
\end{tcolorbox}
\end{minipage}

\vspace{1.0pt}

\begin{minipage}[t]{0.333\linewidth}
\begin{tcolorbox}[fixedbox, title=Test generated from bug-free method,
                  equal height group=row2]
\begin{lstlisting}[style=codeTiny, escapeinside={(*@}{@*)}]
@Test
public void testEquals_sb() {
  (*@\hltok{assertTrue}@*)(StringUtils.equals("abc",
   new StringBuilder("abc")));
}(*@\hfill\badgegood{test asserts correct behavior}@*)
\end{lstlisting}
\end{tcolorbox}
\end{minipage}%
\hfill
\begin{minipage}[t]{0.310\linewidth}
\begin{tcolorbox}[buggybox, title=Misguided Test,
                  equal height group=row2]
\begin{lstlisting}[style=codeTiny, escapeinside={(*@}{@*)}]
@Test
public void testEquals_sb() {
  (*@\hltok{assertFalse}@*)(StringUtils.equals("abc",
   new StringBuilder("abc")));
}(*@\hfill\badgebad{test asserts buggy behavior}@*)
\end{lstlisting}
\end{tcolorbox}
\end{minipage}%
\hfill
\begin{minipage}[t]{0.349\linewidth}
\begin{tcolorbox}[fixedbox, title=Test generated from only the docstring,
                  equal height group=row2]
\begin{lstlisting}[style=codeTiny, escapeinside={(*@}{@*)}]
@Test
public void testEquals_sb() {
  (*@\hltok{assertTrue}@*)(StringUtils.equals("abc",
   new StringBuilder("abc")));
}(*@\hfill\badgegood{test asserts correct behavior}@*)
\end{lstlisting}
\end{tcolorbox}
\end{minipage}

\end{minipage}
\vspace{-0.6em}
\caption{Example of a ``misguided test'' generated from buggy code that asserts its buggy behavior, and how the docstring generated from the buggy code can help correct this behavior and produce a test that detects the bug. The top row shows the main input to the LLM, and the bottom row shows the corresponding generated tests. Highlighted text marks the key differences between the code inputs and the generated tests.}
\Description{Side-by-side comparison of the bug-free and buggy implementations of StringUtils.equals, the docstring generated from the buggy code, and the tests generated from each input; the test generated from buggy code asserts the buggy behavior, while the docstring-based test detects the bug.}
\label{fig:misguided_test_example}
\vspace{-0.8em}
\end{figure*}

To address these shortcomings and accurately quantify the misguidance effect, we conduct a large-scale empirical study that leverages a new metric explicitly measuring ``misguided tests,'' defined as tests that pass on the buggy version but fail on its bug-free counterpart. Figure~\ref{fig:misguided_test_example} presents a motivating example of misguided tests using \texttt{StringUtils.equals} from bug Lang-14 in Defects4J. This method is intended to compare the contents of two \texttt{CharSequence} objects, where \texttt{CharSequence} is a Java interface implemented by \texttt{String}, \texttt{StringBuilder}, and \texttt{StringBuffer}; thus \texttt{"abc"} and \texttt{new StringBuilder("abc")} should be considered equal despite their different concrete types. The bug-free implementation handles this case using \texttt{regionMatches}, whereas the buggy implementation directly calls \texttt{cs1.equals(cs2)}. Since \texttt{String.equals} returns \texttt{true} only when its argument is also a \texttt{String} with the same contents, it incorrectly returns \texttt{false} when comparing a \texttt{String} with a \texttt{StringBuilder}. When prompted with this buggy implementation, the LLM treats the faulty behavior as intended and generates a test that asserts this faulty behavior. 

To compute the new metric, we leverage a parallel dataset of real-world buggy code snippets and their fixed versions, allowing us to measure how test results differ for buggy and bug-free code with the same intended functionality. By generating and executing tests against both versions, we reveal a critical twofold impact: relative to the fixed counterpart, prompting LLMs with buggy code significantly increases ``misguided tests'' while suppressing ``effective tests''---tests that successfully identify the bug. Furthermore, our correlation analysis reveals a strong positive correlation between the reduction in misguided tests and the increase in effective tests, suggesting that mitigating the misguidance effect may lead to more effective test suites. To validate the misguidance effect from a model-internal perspective, we also analyze the sequence score, a common metric for measuring an LLM's preference among candidate texts \cite{LMareFew-Shot-Learner, DetectGPT, Holtzman2020TheCuriousCase}, and demonstrate that buggy input code skews this preference toward tests asserting its erroneous behavior. We further show that this detrimental effect extends to more advanced techniques, degrading the multi-round, feedback-driven prompting pipelines common in recent test generation work \cite{EvalImprChatgptUTGenFSE2024, testforge_multi_round, ChatUniTestFSE2024Demo}.
 
Based on our findings, we contend that the prevailing paradigm of prompting an LLM with the code under test has a fundamental limitation: when the code is buggy, its \emph{misguidance effect} can steer the model toward generating tests that assert the very behaviors they are meant to expose. To mitigate this effect, we draw on principles from specification-based testing~\cite{Specification-Base-Testing} (black-box testing), widely adopted in Test-Driven Development (TDD)~\cite{beck2022test-driven-dev}, which derives tests from intended behavior rather than the implementation. We propose a two-step workflow: (1) use the LLM to derive a docstring that captures intended functionality while omitting implementation details; and (2) \emph{replace} the source code in the prompt with this docstring, removing the buggy implementation as a source of misguidance. This design choice is crucial: for buggy inputs, we find that the code must be removed entirely, rather than merely supplemented, to meaningfully mitigate misguidance, departing from prior studies that use LLM-generated documentation only as additional context alongside the code~\cite{EvalImprChatgptUTGenFSE2024}. Figure~\ref{fig:misguided_test_example} also illustrates this approach on the same \texttt{StringUtils.equals} example: when the LLM-generated docstring replaces the buggy code in the test-generation prompt, the generated test correctly asserts that \texttt{"abc"} and \texttt{new StringBuilder("abc")} should be considered equal, thereby exposing the bug. Notably, even a simple specification-construction prompt yields markedly more effective test suites, and we further strengthen this baseline with an analysis-driven intent-derivation step that produces more accurate specifications, which further reduces misguided tests and increases effective tests without substantially increasing tests that assert hallucinated behavior present in neither the buggy nor the fixed code. Our specification-based approach also enhances the performance of the interactive, multi-round test-generation pipelines adopted in recent work~\cite{ChatUniTestFSE2024Demo, EvalImprChatgptUTGenFSE2024, testforge_multi_round}. We also manually inspect the quality of the generated docstrings and quantify its effect on the resulting tests, showing that our approach effectively blocks bug propagation into generated docstrings and recovers the correct behavior for a considerable fraction of buggy focal methods. The former results in significantly fewer misguided test suites, while the latter contributes to substantially more detected bugs, supporting our design intuition and underscoring the need for methods that can reconstruct more accurate specifications from buggy code. Overall, these results confirm that our specification-based approach is effective at mitigating the misguidance effect.

Finally, since it is often unknown whether the code under test is indeed buggy, we demonstrate that our approach can be applied to both buggy and bug-free code. On bug-free code, it produces comparable levels of compilation failures, false-alarm tests (tests that fail on correct code), and test coverage relative to the baseline of using the source code as input, creating no additional burden for developers while keeping the generated tests applicable to broader uses like regression testing.

In summary, this paper makes the following contributions:

\newcommand{\myitem}[1]{%
\par\noindent
\hangindent=10pt
\hangafter=1
\makebox[10pt][l]{\labelitemi}#1\par
}

\myitem{We conduct a large-scale quantitative study that leverages a new metric to empirically validate the \emph{Misguidance Effect} of buggy code and confirms this effect from a model-internal perspective. Our correlation analysis further reveals a strong positive correlation between mitigating the misguidance effect and improving the effectiveness of the generated unit tests.}

\myitem{We introduce and evaluate a \emph{specification-based} test generation paradigm that uses an LLM-generated behavioral specification as the core of the prompt, rather than the buggy code. We show that this paradigm significantly reduces misguided tests while improving bug-detection rates. Furthermore, we show that it enhances the efficacy of an interactive, multi-round prompting pipeline and analyze how the quality of the generated specification affects the final test suite.}

\myitem{We demonstrate the robustness of our method for both buggy and bug-free code. When applied to bug-free code, it yields comparable levels of compilation failures, false-alarm tests, and test coverage to the baseline. This ensures that our method enhances bug detection without adding extra burden to developers or compromising its utility for broader uses like regression testing.}

\section{Study Design}  
\subsection{Research Questions}
\label{subsec:research_questions}
For our study, we investigated the following research questions (RQs) to identify, analyze, and mitigate the misguidance effect and its negative impact on bug detection.
\begin{itemize}
    \item \textbf{RQ1}: How does the misguidance effect of buggy code affect the behavior of LLM-generated unit tests?
    \item \textbf{RQ2}: How does our proposed specification-based unit test generation pipeline help mitigate the misguidance effect?
    \item \textbf{RQ3}: Does our proposed specification-based unit test generation pipeline impact unit tests generated from bug-free code?
\end{itemize}
\noindent\textbf{RQ1} investigates the existence and severity of the misguidance effect. We conduct a large-scale evaluation to provide empirical evidence of this phenomenon, verify its detrimental impact on the bug detection capability of the generated tests, and confirm this effect from a model-internal perspective. Furthermore, we analyze the correlation between mitigating this effect and improving bug detection effectiveness.

\noindent\textbf{RQ2} investigates the efficacy of our specification-based pipeline on buggy code. We first evaluate its ability to mitigate misguidance and improve bug detection, then examine whether its gains scale with specification quality and persist within multi-round, feedback-driven prompting pipelines. Finally, we manually inspect whether bugs are inherited by the LLM-generated specifications and how this affects the resulting test suites.

\noindent\textbf{RQ3} evaluates the impact of our specification-based pipeline on bug-free code to confirm its applicability to both buggy and bug-free code, ensuring it does not impose unnecessary burdens on developers or compromise its utility for broader applications like regression testing.

\subsection{Benchmark}

In this study, we employ Defects4J \cite{Defects4j}, a widely used dataset containing real bugs from open-source Java projects (e.g., JFreeChart, Commons Lang). Version 3.0 of the dataset includes 854 defects (bugs) across 17 projects. Each defect comes with both a buggy and a fixed version of the code and may contain one or more buggy methods, which we refer to as focal methods. Human-written tests, including those that detect the buggy focal methods, are also provided.

Following prior work on evaluating LLM-based unit test generation \cite{OnLLMUTGenEvalASE2024, EvalImprChatgptUTGenFSE2024, EmpEvalforLLMUTAutoGenTSE}, we focus test generation and evaluation on focal methods. To select focal methods, we follow the general practice of prior studies \cite{OnLLMUTGenEvalASE2024, A3TestIST, UnitTestCaseGenTrans} and further refine the selection based on the following criteria:
(1) We include all non-private methods (public, protected, and default) in our analysis, as their accessibility from the test package makes them directly testable.
(2) The focal method must exist in both the buggy and fixed versions to enable meaningful performance comparisons. Focal methods that are removed or newly introduced by the patch are excluded.
(3) We only retain focal methods that trigger at least one existing human-written test in Defects4J with its buggy version, ensuring that the corresponding patch is for bug fixes rather than stylistic or performance optimizations. In total, our curated benchmark comprises 318 focal methods covering 233 defects across all 17 projects.




\subsection{Test Generation Workflow}
\label{subsec:test_gen_workflow}
To support a realistic and generalizable evaluation, we adopt a test-generation workflow that closely follows the end-to-end pipeline for evaluating LLM-generated unit tests proposed by Yang et al.~\cite{OnLLMUTGenEvalASE2024}, spanning prompt construction and unit test extraction. Following their findings on effective prompt context, we augment the focal method with additional, readily obtainable surrounding code features, including the method signature and parameters, the enclosing class constructor, declared fields, and other methods in the same class. Furthermore, we include constructors of user-defined classes that appear as parameter types or return types, giving the LLM the information needed to instantiate these object dependencies and generate more reliable tests. Depending on the experimental setting, the focal method body and/or a docstring generated from the focal method is used as the core behavioral input to the prompt. We wrap this context with a system instruction that positions the LLM as a professional Java developer and append an explicit request to generate unit tests for the provided code. We include a summary of the test generation prompt in Figure~\ref{fig:prompt_template}.

\begin{figure}[htbp!]
\vspace{-0.8em}
\begin{minipage}{0.98\linewidth}
\begin{lstlisting}[style=promptstyle]
You are a professional who writes Java test methods. Please help me write some unit tests in Java language, details are listed below: 
```
<Code under test and/or LLM-generated docstring, and other focal method-related contexts>
```
Please write some unit tests in Java 11 and Junit 4 with maximizing both branch and line coverage. Please ensure that the output format is Markdown, and no explanations needed.
\end{lstlisting}
\end{minipage}
\vspace{-0.8em}
\caption{The prompt template used for generating unit tests.}
\Description{The prompt template used for generating unit tests, consisting of a system instruction, the code under test or docstring with related context, and an explicit generation request.}
\vspace{-0.8em}
\label{fig:prompt_template}
\end{figure}

We adopt Yang et al.'s prompt configuration for three reasons. First, they evaluate a broad set of open- and closed-source models, supporting the generality of the prompt design. Second, their ablation study systematically analyzes the impact of additional code context and motivates the specific context features we include. Third, the required context can be extracted solely from the code under test without manual inspection, enabling a fully automated prompt-construction and test-generation pipeline that aligns with the practical goal of reducing developer burden.

To extract unit tests from LLM outputs, we use an abstract syntax tree (AST) parser (Tree-sitter~\cite{tree-sitter}) to identify generated test cases and related artifacts such as imports and helper functions. We then compose the final test files by combining the extracted content with project-specific dependencies and a curated set of common dependencies for the JDK (e.g., \texttt{java.util}) and JUnit (e.g., \texttt{org.junit.Assert}), to prevent compilation failures from missing imports.

\subsection{Models}
\label{subsec:models}
To ensure a comprehensive analysis, we selected a diverse set of 11 state-of-the-art (SOTA) Large Language Models from six leading developers, including both ``base'' and ``reasoning'' models where available. Reasoning models are optimized for multi-step reasoning (often via an explicit chain-of-thought) to improve performance on tasks requiring complex logical deduction, while base models generate output directly. For hybrid models capable of operating in either mode, such as Claude 4 Sonnet \cite{anthropic2025claude4sonnet}, we evaluated their performance with and without reasoning enabled. A full list of the evaluated models is presented in Table~\ref{tab:evaluated_models}.

\begin{table}[htbp!]
\vspace{-0.6em}
\centering
\scriptsize
\caption{LLMs evaluated for unit test generation in our study.}
\vspace{-0.6em}
\label{tab:evaluated_models}
\begin{tabular}{lllc}
\toprule
\textbf{Developer} & \textbf{Model Name} & \textbf{Type} & \textbf{Open Sourced} \\
\midrule
\multirow{2}{*}{Google} & \emph{Gemini 2.5 Pro} \cite{Google2025Gemini2.5Pro} & Reasoning & \ding{55} \\
 & \emph{Gemini 2.5 Flash} \cite{Google2025Gemini2.5Flash} & Hybrid & \ding{55} \\
\cmidrule(lr){1-4}
Anthropic & \emph{Claude 4 Sonnet} \cite{anthropic2025claude4sonnet} & Hybrid & \ding{55} \\
\cmidrule(lr){1-4}
\multirow{2}{*}{xAI} & \emph{Grok-4} \cite{xai2025grok4} & Reasoning & \ding{55} \\
 & \emph{Grok-3} \cite{xai2025grok3} & Base & \ding{55} \\
\cmidrule(lr){1-4}
\multirow{2}{*}{OpenAI} & \emph{GPT-4.1} \cite{openai2025gpt41} & Base & \ding{55} \\
 & \emph{GPT-O4-mini} \cite{openai2025o4mini} & Reasoning & \ding{55} \\
\cmidrule(lr){1-4}
\multirow{2}{*}{DeepSeek} & \emph{DeepSeek-V3} \cite{deepseekai2025deepseekv3technicalreport} & Base & \ding{51} \\
 & \emph{DeepSeek-R1} \cite{deepseekai2025deepseekr1incentivizingreasoningcapability} & Reasoning & \ding{51} \\
\cmidrule(lr){1-4}
\multirow{2}{*}{Alibaba} & \emph{Qwen3-Coder-Plus} \cite{alibaba2025qwen3coder} & Base & \ding{51} \\
 & \emph{Qwen3-Plus} \cite{alibaba2025qwen3} & Reasoning & \ding{51} \\
\bottomrule
\end{tabular}
\vspace{-1.5em}
\end{table}

\begin{figure*}[b]
\centering
\setlength{\tabcolsep}{4pt}
\renewcommand{\arraystretch}{1.08}
\vspace{-0.5em}
\begin{tabularx}{\dimexpr\textwidth-8pt\relax}{@{}LZ@{}}
\toprule
\rowcolor{PromptLabel}
Prompt component & Prompt text \\
\midrule

\cellcolor{PromptShared}\promptcomponent{Shared Prompt Main Body}{S.P.M.B.}
&
\cellcolor{PromptShared}
You are a professional Java developer. Please help identify the intention and functionality of the method detailed below:

\promptcode{<code under test and related information>}
\\

\addlinespace[2pt]
\cellcolor{PromptBase}\promptcomponent{Base Docstring Prompt Postfix}{B.D.P.P.}
&
\cellcolor{PromptBase}
Please provide a formal docstring that identifies the functionality and intention of the given method. Please avoid directly quoting from the focal method code, as it might be buggy, and output only the docstring.
\\

\addlinespace[2pt]
\cellcolor{PromptAdvanced}\promptcomponent{Advanced Analysis Prompt}{A.A.P.}
&
\cellcolor{PromptAdvanced}
Please provide a formal docstring that identifies the functionality and intended specification of the given method. To do this, first perform a critical analysis from the perspective of an expert software engineer auditing for quality and correctness. Your analysis must identify two types of potential issues:

\textbf{1. Logical Mistakes:} Scrutinize the algorithm's logic. Based on the method's name and context, does its implementation correctly achieve its apparent goal, or are there logical flaws that would produce an incorrect result even on a ``happy path''?

\textbf{2. Robustness Omissions:} Check for missing but necessary steps that production-quality code would include. This includes, but is not limited to, input validation, e.g., null checks and boundary conditions, proper error handling, and necessary data sanitization or escaping.
\\

\addlinespace[2pt]
\cellcolor{PromptReasoning}\promptcomponent{Output Requirement for Reasoning Models}{O.R.R.M.}
&
\cellcolor{PromptReasoning}
Make sure to clearly write out your analysis. Based on this two-part analysis, the docstring should describe the specification for a correct and robust version of the method, capturing the developer's likely intent. Please avoid directly quoting from the focal method code, as it might be buggy. Output the docstring in a fenced Java Markdown block, between \promptcode{\javafence} and \promptcode{\mdfence}, with the docstring itself wrapped between \promptcode{/**} and \promptcode{*/}.
\\

\addlinespace[2pt]
\cellcolor{PromptBaseModel}\promptcomponent{Output Requirement for Base Models}{O.R.B.M.}
&
\cellcolor{PromptBaseModel}
Your output should come in three parts.

\textbf{Part 1: Critical Analysis.} Clearly state the logical mistakes and robustness omissions found. If no issues are found, explicitly state: ``No issues found.''

\textbf{Part 2: Fixes Required.} List all fixes required to correct the identified logical errors and robustness omissions. If no fixes are required, state: ``No fixes required.''

\textbf{Part 3: Final Specification Docstring.} Based on Part 2, write a formal docstring describing the corrected behavior of the method with all proposed fixes applied. Please avoid directly quoting from the focal method code, as it might be buggy. Output the docstring in a fenced Java Markdown block, between \promptcode{\javafence} and \promptcode{\mdfence}, with the docstring itself wrapped between \promptcode{/**} and \promptcode{*/}.
\\

\addlinespace[2pt]
\cellcolor{PromptLabel}\textsc{Prompt Assembly}
&
\cellcolor{PromptLabel}
\textbf{Base Docstring Prompt} =
\prompttag{PromptShared}{S.P.M.B.} + \prompttag{PromptBase}{B.D.P.P.}

\textbf{Advanced Prompt for Reasoning Models} =
\prompttag{PromptShared}{S.P.M.B.} + \prompttag{PromptAdvanced}{A.A.P.} + \prompttag{PromptReasoning}{O.R.R.M.}

\textbf{Advanced Prompt for Base Models} =
\prompttag{PromptShared}{S.P.M.B.} + \prompttag{PromptAdvanced}{A.A.P.} + \prompttag{PromptBaseModel}{O.R.B.M.}
\\

\bottomrule
\end{tabularx}

\vspace{-0.3em}
\caption{Merged prompt templates for generating specification docstrings from code under test. Each row shows a reusable prompt component; the final row specifies how the three prompt variants are assembled.}
\Description{Table of reusable prompt components for generating specification docstrings, with the assembly rules for the base and advanced prompt variants.}
\label{fig:docstring_prompt_templates}
\vspace{-0.5em}
\end{figure*}

\subsection{Mitigation Methodology and Baselines}
\label{subsec:2.5_methodology}
To counteract the misguidance effect caused by buggy code, we propose a specification-based testing approach. Prior studies~\cite{huang2025measuringinfluenceincorrectcode, EmpEvalforLLMUTAutoGenTSE, trickybug} have explored using high-quality, human-written documentation to improve LLM-generated tests, but such documentation is often unavailable in practice. Instead, we use the LLM to recover the intended behavior of the code under test and focus test generation on it rather than on the potentially erroneous implementation details.

Our approach consists of a two-step process. First, we prompt the LLM to construct a behavioral specification of the code in the form of a docstring. Second, we use this specification docstring as the only behavioral input for the test-generation model, discarding the original buggy implementation. By compelling the model to operate on the specification rather than the buggy code, we aim to reduce its bias toward implementation flaws, thereby reducing the number of misguided tests while increasing the number of effective bug-finding tests. Compared with direct code-based test generation, the main additional overhead of our approach is one extra LLM call for docstring generation and the associated input/output token cost. This represents a key departure from prior work, which either uses buggy code as the sole input to the LLM~\cite{EmpEvalforLLMUTAutoGenTSE, ChatUniTestFSE2024Demo, CODAMOSAICSE2023} or uses LLM-generated documentation only as supplementary context alongside the buggy code~\cite{EvalImprChatgptUTGenFSE2024}. We compare against these settings as baselines, as well as a baseline that removes both the code under test and the generated docstring, to assess whether both components of our design are necessary: (1) fully removing the code under test, and (2) replacing it with an LLM-generated specification. We present the prompt for the base version of our approach in Figure~\ref{fig:docstring_prompt_templates}, labeled \textsc{Base Docstring Prompt}.

To further strengthen our approach, we introduce an advanced prompting strategy (the \textsc{Advanced Docstring Prompt}) that leverages modern LLMs' code-comprehension capabilities by requiring the model to perform a two-part analysis before generating the docstring: (1) identify logical errors where the implementation contradicts the intent inferred from the method name and context, and (2) detect robustness gaps, such as missing null checks, inadequate sanitization, or unhandled edge cases. We present the prompt that enforces this analysis in the \textsc{Advanced Analysis Prompt} row of Figure~\ref{fig:docstring_prompt_templates}. This strategy is implemented differently for reasoning and base models to maximize its effectiveness. For reasoning models, asking the model to analyze these issues is sufficient to elicit the necessary reasoning process. For base models, however, the same instruction does not reliably elicit the required reasoning steps. We therefore require the model to explicitly report the identified bugs and the fixes needed to address them, and only then write the docstring under the assumption that those fixes have been applied. We evaluate this advanced prompt against two baselines: (1) the \textsc{Base Docstring Prompt}, to measure the benefit of advanced analysis over the base version of our approach, and (2) direct code-based test generation with the same advanced-analysis instruction. This second baseline generates tests in a single step without \emph{removing the buggy implementation} or \emph{replacing it with an LLM-generated docstring}. This comparison tests whether advanced analysis alone can mitigate the misguidance effect when the buggy code under test remains visible, or whether it must be combined with our specification-based approach to generate a docstring from the code under test that replaces the code during test generation.

Finally, we manually inspect all 636 docstrings generated with the \textsc{Advanced Docstring Prompt} by the two models that achieved the largest and smallest reductions in misguided test suites after applying our approach. We use this inspection to analyze how docstring quality affects downstream test generation. We detail the inspected docstring properties in the next section.

\subsection{Metrics and Statistics}
\begin{table}[!htbp]
    \vspace{-0.5em}
    \centering
    \scriptsize
    \caption{Test categories based on outcomes across two code versions. ``Positive'' indicates a flagged bug, while ``Negative'' means no bug was flagged.}
    \vspace{-0.4em}
    \label{tab:err_cats}
    \begin{tabular}{l|cc}
    \toprule
    \textbf{Categories} & \textbf{Fixed} & \textbf{Buggy}  \\
\midrule
     True Negative    & Passed & Passed \\
     True Positive (Effective)   & Passed & Failed \\
     False Negative (Misguided)   & Failed & Passed \\
     False Positive   & Failed & Failed \\
     \bottomrule
    \end{tabular}
\vspace{-0.6em}
\end{table}
\label{subsec:metrics}

\begin{table}[ht]
\vspace{-0.4em}
\centering
\scriptsize
\caption{Tests that pass on the buggy version, with the counts (\#) and ratios (\%) of \emph{True Negative} (pass on both) and \emph{False Negative} (misguided) tests among them.}
\vspace{-0.4em}
\label{tab:flaw_demo}
\begin{tabular}{l|c|cc|cc}
\toprule
\multirow{3}{*}{\textbf{Model}} & \multirow{2}{*}{\textbf{Passed on Buggy}} & \multicolumn{2}{c|}{\textbf{True Negative Tests}} & \multicolumn{2}{c}{\textbf{False Negative Tests}} \\
\cmidrule(lr){3-4}\cmidrule(lr){5-6}
 &  & \textbf{\#} & \textbf{\%} & \textbf{\#} & \textbf{\%} \\
\midrule
Gemini 2.5 Pro & 2183 & 2010 & 92.08\% & 173 & 7.92\% \\
Gemini 2.5 Flash & 1782 & 1668 & 93.60\% & 114 & 6.40\% \\
Gemini 2.5 Flash (Reason) & 2066 & 1911 & 92.50\% & 155 & 7.50\% \\
Claude 4 Sonnet & 3034 & 2885 & 95.09\% & 149 & 4.91\% \\
Claude 4 Sonnet (Reason) & 2932 & 2745 & 93.62\% & 187 & 6.38\% \\
Grok-4  & 2482 & 2303 & 92.79\% & 179 & 7.21\% \\
Grok-3 & 1468 & 1397 & 95.16\% & 71 & 4.84\% \\
GPT-4.1 & 1806 & 1714 & 94.91\% & 92 & 5.09\% \\
GPT-O4-mini & 1672 & 1542 & 92.22\% & 130 & 7.78\% \\
DeepSeek-V3 & 1750 & 1674 & 95.66\% & 76 & 4.34\% \\
DeepSeek-R1 & 2206 & 2061 & 93.43\% & 145 & 6.57\% \\
Qwen3-Coder-Plus & 2053 & 1973 & 96.10\% & 80 & 3.90\% \\
Qwen3-Plus & 2914 & 2675 & 91.80\% & 239 & 8.20\% \\
\midrule
Average & 2181 & 2043 & 93.77\% & 138 & 6.23\% \\
\bottomrule
\end{tabular}
\vspace{-1.8em}
\end{table}

To address our research questions, we adopt several key metrics derived from executing each generated test against both the buggy program and its corresponding fixed version. The tests are classified into four categories based on the outcomes of this dual execution, as detailed in Table~\ref{tab:err_cats}.

Throughout our evaluation, we treat the fixed Defects4J version as the gold standard for intended program behavior and define all test labels accordingly. Under this definition, any implementation that produces externally observable outputs that differ from the fixed version for the same inputs is not considered bug-free in our evaluation. This benchmark-specific oracle allows us to consistently label misguided, effective, and false-alarm tests across all focal methods.

For RQ1, to measure the misguidance effect and its impact on bug detection effectiveness, we compare the number and proportion of two types of tests generated from buggy versus fixed code:
\begin{itemize}
\item \textbf{Misguided tests} (False Negatives): tests that pass on the buggy code and fail on its corresponding fixed version, which serve to validate and quantify the misguidance effect. 
\item \textbf{Effective tests} (True Positives): tests that fail on the buggy code and pass on its corresponding fixed version, which directly measure the impact on bug detection capability.
\end{itemize}

Our metric for quantifying the misguidance effect differs substantially from Huang et al.'s \cite{huang2025measuringinfluenceincorrectcode}. We count as ``misguided'' only False Negatives—i.e., tests that pass on the buggy version but fail on the fixed version—because only these tests explicitly assert the buggy behavior. In contrast, Huang et al. treat all tests that pass on the buggy version (True Negatives + False Negatives) as evidence of misguidance; however, this aggregated count can be driven primarily by True Negatives—valid tests that pass on both versions. To illustrate this limitation, we compute and report in Table~\ref{tab:flaw_demo} the counts and ratios of True Negatives (pass on both) and False Negatives (pass only on buggy). We find that, among tests that pass on buggy code, the vast majority—over 90\% on average—are, in fact, True Negatives. Consequently, Huang et al.'s metric based on the aggregate count does not accurately capture the existence and extent of the misguidance effect of buggy code. 

To further validate these behavioral findings of LLM-generated tests from a model-internal perspective, we analyze the sequence score \cite{LMareFew-Shot-Learner, DetectGPT, Holtzman2020TheCuriousCase} that an LLM assigns to a given test (misguided or effective) when conditioned on buggy or fixed source code. This score is formally defined as:
\begin{equation}
\label{eq:normalized_sequence_score}
S(T, C) = \frac{1}{|T|} \sum_{i=1}^{|T|} \log P(t_i | C, t_1, \ldots, t_{i-1})
\end{equation}
In this equation, $S(T, C)$ represents the normalized score of a generated test sequence $T$ given input code context $C$. This score is calculated as the average log probability per token, which reflects the model's confidence in generating that specific sequence. Normalizing by the sequence length $|T|$ ensures a fair comparison between tests with different token sequence lengths \cite{LMareFew-Shot-Learner}. 

For RQ2, we measure changes in both misguided and effective tests to confirm that our approach mitigates misguidance and improves bug detection. To ensure gains are not concentrated in a few focal methods, we also analyze results at the method level, reporting the number of unique methods where the LLM is misguided (at least one misguided test) and the number of unique buggy focal methods detected (at least one effective test). This verifies effectiveness across the dataset at the focal-method granularity. 
For the advanced analysis approach, because the model may hallucinate a non-existent ``false-intent'' or incorrect fix, we also report tests that assert hallucinated behavior present in neither the buggy nor the fixed version, i.e., \emph{False Positive} tests. We further conduct a manual inspection to annotate whether each LLM-generated docstring (1) preserves the original bug from the buggy focal method, (2) describes the corrected behavior needed to fix the bug.

For RQ3, we measure test correctness metrics~\cite{OnLLMUTGenEvalASE2024, EvalImprChatgptUTGenFSE2024, EmpEvalforLLMUTAutoGenTSE}, including the rates of compilation failures and false-alarm tests. Figure~\ref{fig:false-alarm-example} presents an example of a false-alarm test, i.e., a test that fails on bug-free code by asserting hallucinated behavior not present in that code. We also measure test coverage metrics~\cite{HITS:HighCoveLLMUTGen, CovNotStronglyCorrWithEffICSE2014, CAT-LMASE2023}, including line and branch coverage for the bug-free code.

\begin{figure*}[htbp!]
\vspace{-0.8em}
\begin{minipage}{1.00\textwidth}%
\centering

\begin{minipage}[t]{0.498\linewidth}%
\vspace{0pt}%
\begin{tcolorbox}[fixedbox,
    equal height group=falsealarm-method-only,
    title=Bug-free focal method]
\begin{lstlisting}[style=codeTiny]
public String generateToolTipFragment(String toolTipText) {
  return " title=\"" + ImageMapUtilities.htmlEscape(toolTipText)
     + "\" alt=\"\"";
}
\end{lstlisting}
\end{tcolorbox}%
\par\vspace{-5.0pt}%
\begin{tcolorbox}[buggybox,
    title=Test asserting the hallucinated behavior]
\begin{lstlisting}[style=codeTiny,
    linebackgroundcolor={\hlRange{4}{4}}]
@Test
public void testGenerateToolTipFragment_Null() {
  String result = gen.generateToolTipFragment(null);
  Assert.assertEquals("", result);
}
\end{lstlisting}
\end{tcolorbox}%
\end{minipage}%
\hfill%
\begin{minipage}[t]{0.493\linewidth}%
\vspace{0pt}%
\begin{tcolorbox}[docstringbox,
    title=Generated docstring with hallucinated behavior]
\begin{lstlisting}[style=codeTiny,
    linebackgroundcolor={\hlRange{4}{6}\hlRange{10}{11}}]
/**
 * Generates a sanitized HTML attribute fragment...
 * <ul>
 *   <li>Gracefully handle null... treating null as an empty
 *       string rather than producing a literal "null" string
 *       or throwing an exception.</li>
 *   <li>Safely escape all special characters...</li>
 * </ul>
 *
 * @return a valid HTML attribute fragment containing the
 *         escaped title, or an empty string if input is null
 */
\end{lstlisting}
\end{tcolorbox}%
\end{minipage}%

\end{minipage}

\vspace{-0.6em}
\caption{Example of a false-alarm test generated from a hallucinated docstring. The highlighted lines mark the hallucinated behavior in the generated docstring and the corresponding erroneous assertion in the test.}
\Description{Example of a hallucinated docstring describing behavior present in neither code version, and the resulting false-alarm test that fails on both versions.}
\vspace{-1.5em}
\label{fig:false-alarm-example}
\end{figure*}

\section{RQ1: Misguidance Effect of Buggy Code}
\label{sec:rq1_misguidance}
\begin{figure}[htbp!]
    \vspace{-0.5em}
    \centering
    \includegraphics[width=\linewidth]{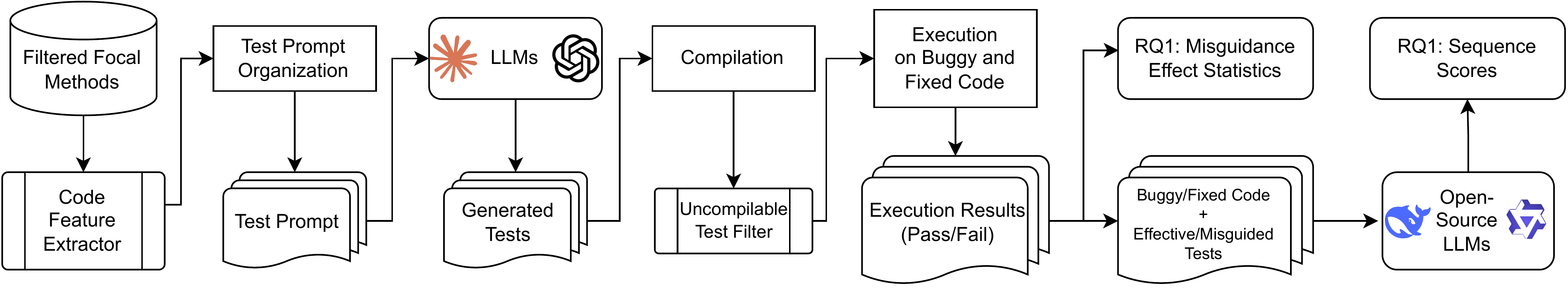}
    \vspace{-0.6em}
    \caption{Workflow for evaluation of the misguidance effect from buggy code.}
    \Description{Workflow diagram: tests are generated from the buggy and fixed versions of each focal method, executed against both versions, and classified to measure the misguidance effect.}
    \vspace{-1.0em}
    \label{fig:eval_workflow}
\end{figure}

In this section, we evaluate the misguidance effect of buggy code and its impact on the bug detection effectiveness of generated tests from both test performance and model-internal perspectives. Our evaluation workflow is presented in Figure~\ref{fig:eval_workflow}. 

\subsection{The Misguidance Effect on Test Generation}
\label{subsec:misguidance_effect_statistics}

Table~\ref{tab:misguidance_effect} compares tests generated from buggy versus fixed code. When prompted with buggy code, models produce substantially more misguided tests, which incorrectly validate buggy behavior by passing on the buggy code but failing on the fixed version. On average, the models generate 137.69 misguided tests (3.84\%). In contrast, when given the corresponding fixed code, the average drops to 16.46 (0.46\%). This gap provides quantitative evidence of the existence and magnitude of the misguidance effect, suggesting that modern LLMs can misinterpret erroneous logic as intended functionality.

\begin{table}[htbp!]
\vspace{-0.5em}
\centering
\tiny
\caption{Total tests generated per model, with the number (\#) and percentage (\%) of misguided and effective tests for buggy and fixed code inputs, grouped by base and reasoning model types.}
\vspace{-0.4em}
\label{tab:misguidance_effect}
\begin{tabular}{l | c | cc | cc | c | cc | cc}
\toprule
\multirow{4}{*}{\textbf{Model}} & \multicolumn{5}{c|}{\textbf{Buggy Input}} & \multicolumn{5}{c}{\textbf{Fixed Input}} \\
\cmidrule{2-11}
& \multirow{2}{*}{\textbf{Total}} & \multicolumn{2}{c|}{\emph{Misguided}} & \multicolumn{2}{c|}{\emph{Effective}} & \multirow{2}{*}{\textbf{Total}} & \multicolumn{2}{c|}{\emph{Misguided}} & \multicolumn{2}{c}{\emph{Effective}}\\
\cmidrule(lr){3-4} \cmidrule(lr){5-6} \cmidrule(lr){8-9} \cmidrule(lr){10-11}
& & \textbf{\#} & \textbf{\%} & \textbf{\#} & \textbf{\%} & & \textbf{\#} & \textbf{\%} & \textbf{\#} & \textbf{\%} \\
\midrule
Gemini 2.5 Flash & 3539 & 114 & 3.22 & 140 & 3.96 & 3616 & 11 & 0.30 & 279 & 7.72 \\
Claude 4 Sonnet & 4202 & 149 & 3.55 & 88 & 2.09 & 4250 & 15 & 0.35 & 361 & 8.49 \\
Grok-3 & 2543 & 71 & 2.79 & 93 & 3.66 & 2581 & 17 & 0.66 & 199 & 7.71 \\
GPT-4.1 & 2841 & 92 & 3.24 & 86 & 3.03 & 2905 & 11 & 0.38 & 236 & 8.12 \\
DeepSeek-V3 & 3002 & 76 & 2.53 & 113 & 3.76 & 3008 & 19 & 0.63 & 241 & 8.01 \\
Qwen3-Coder-Plus & 3455 & 80 & 2.32 & 157 & 4.54 & 3353 & 10 & 0.30 & 259 & 7.72 \\
\midrule
Base Average & 3263.67 & 97.00 & 2.94 & 112.83 & 3.51 & 3285.50 & 13.83 & 0.44 & 262.50 & 7.96 \\
\midrule
Gemini 2.5 Pro & 3284 & 173 & 5.27 & 98 & 2.98 & 3286 & 17 & 0.52 & 354 & 10.77 \\
Gemini 2.5 Flash (Reason) & 3957 & 155 & 3.92 & 99 & 2.50 & 3956 & 7 & 0.18 & 293 & 7.41 \\
Claude 4 Sonnet (Reason) & 4515 & 187 & 4.14 & 88 & 1.95 & 4553 & 27 & 0.59 & 372 & 8.17 \\
Grok-4 & 3923 & 179 & 4.56 & 145 & 3.70 & 3985 & 19 & 0.48 & 398 & 9.99 \\
GPT-O4-mini & 2500 & 130 & 5.20 & 36 & 1.44 & 2606 & 7 & 0.27 & 248 & 9.52 \\
DeepSeek-R1 & 3413 & 145 & 4.25 & 97 & 2.84 & 3622 & 21 & 0.58 & 324 & 8.95 \\
Qwen3-Plus & 4857 & 239 & 4.92 & 114 & 2.35 & 4808 & 33 & 0.69 & 389 & 8.09 \\
\midrule
Reasoning Average & 3778.43 & 172.57 & 4.61 & 96.71 & 2.54 & 3830.86 & 18.71 & 0.47 & 339.71 & 8.99 \\
\midrule
Total Average & 3540.85 & 137.69 & 3.84 & 104.15 & 2.98 & 3579.15 & 16.46 & 0.46 & 304.08 & 8.51 \\
\bottomrule
\end{tabular}
\end{table}

Beyond increasing misguided tests, the misguidance effect also directly undermines bug detection. For effective tests, which fail on the buggy code and pass on the fixed version, models produce only 104.15 on average (2.98\%) when prompted with buggy code. When given the corresponding fixed code, this rises to 304.08 (8.51\%), nearly a threefold increase. These results indicate that buggy input code not only leads to more incorrect tests, but also suppresses the generation of useful, bug-finding ones.
\begin{table}[b]
\vspace{-0.5em}
\centering
\setlength{\tabcolsep}{2.5pt}
\tiny
\caption{Average sequence scores for generated tests under varied input conditions. Rows indicate the test-generating model, and columns indicate the evaluator model. To prevent self-evaluation bias, scores where the generator and evaluator are from the same developer have been excluded. \emph{M}/\emph{E}: misguided/effective tests.}
\vspace{-0.4em}
\label{tab:sequence_score}
\begin{tabular}{l | c c | c c | c c | c c | c c | c c}
\toprule
\multirow{4}{*}{\textbf{Model}} & \multicolumn{4}{c|}{\textbf{DeepSeek-V3}} & \multicolumn{4}{c|}{\textbf{Qwen3-Coder-Plus}} & \multicolumn{4}{c}{\textbf{GPT-OSS-120B}}\\
\cmidrule{2-13}
& \multicolumn{2}{c|}{\textbf{Buggy Input}} & \multicolumn{2}{c|}{\textbf{Fixed Input}} & \multicolumn{2}{c|}{\textbf{Buggy Input}} & \multicolumn{2}{c|}{\textbf{Fixed Input}} & \multicolumn{2}{c|}{\textbf{Buggy Input}} & \multicolumn{2}{c}{\textbf{Fixed Input}} \\
& \emph{M} & \emph{E} &  \emph{M} & \emph{E} & \emph{M} & \emph{E} &  \emph{M} & \emph{E} & \emph{M} & \emph{E} &  \emph{M} & \emph{E}\\
\midrule
Gemini 2.5 Pro & -1.28 & -1.36 & -1.35 & -1.29 & -1.64 & -1.77 & -1.76 & -1.68 & -2.21 & -2.42 & -2.30 & -2.35 \\
Gemini 2.5 Flash & -1.12 & -1.23 & -1.23 & -1.17 & -1.39 & -1.61 & -1.55 & -1.52 & -2.00 & -2.27 & -2.14 & -2.19 \\
Gemini 2.5 Flash (Reason) & -1.24 & -1.38 & -1.31 & -1.32 & -1.61 & -1.79 & -1.73 & -1.71 & -2.14 & -2.38 & -2.23 & -2.31 \\
Claude 4 Sonnet & -1.07 & -1.22 & -1.20 & -1.14 & -1.39 & -1.65 & -1.56 & -1.51 & -2.16 & -2.44 & -2.33 & -2.36 \\
Claude 4 Sonnet (Reason) & -1.10 & -1.26 & -1.21 & -1.19 & -1.44 & -1.69 & -1.62 & -1.58 & -2.17 & -2.46 & -2.34 & -2.39 \\
Grok-4 & -1.38 & -1.37 & -1.45 & -1.29 & -1.85 & -1.85 & -1.99 & -1.71 & -2.68 & -2.58 & -2.80 & -2.50 \\
Grok-3 & -1.23 & -1.29 & -1.35 & -1.23 & -1.58 & -1.73 & -1.75 & -1.61 & -2.44 & -2.51 & -2.59 & -2.43 \\
GPT-4.1 & -1.16 & -1.31 & -1.28 & -1.23 & -1.57 & -1.73 & -1.78 & -1.62 & - & - & - & - \\
GPT-O4-mini & -1.40 & -1.39 & -1.48 & -1.32 & -1.85 & -1.84 & -1.99 & -1.73 & - & - & - & - \\
DeepSeek-V3 & - & - & - & - & -1.49 & -1.70 & -1.72 & -1.57 & -2.28 & -2.61 & -2.46 & -2.53 \\
DeepSeek-R1 & - & - & - & - & -1.70 & -1.87 & -1.84 & -1.75 & -2.49 & -2.70 & -2.60 & -2.61 \\
Qwen3-Coder-Plus & -1.12 & -1.23 & -1.27 & -1.17 & - & - & - & - & -2.27 & -2.52 & -2.45 & -2.44 \\
Qwen3-Plus & -1.21 & -1.29 & -1.28 & -1.22 & - & - & - & - & -2.40 & -2.56 & -2.52 & -2.47 \\
\midrule
Average & -1.21 & -1.30 & -1.31 & -1.23 & -1.59 & -1.75 & -1.75 & -1.64 & -2.29 & -2.50 & -2.43 & -2.42 \\
\bottomrule
\end{tabular}
\vspace{-0.5em}
\end{table}
We also observe that reasoning models tend to generate more misguided tests from buggy input and more effective tests from fixed input. This finding is supported by a correlation analysis. We find a strong positive correlation between the number of misguided tests from buggy input and the number of effective tests from fixed input ($r=0.90, p < 0.01$), as well as a strong correlation between their respective ratios ($r=0.72, p < 0.01$). These correlations suggest that the models most capable of generating correct tests are also the ones most susceptible to being misguided by buggy code. Furthermore, we observe a strong positive correlation between the reduction in misguided tests and the increase in effective tests when switching from buggy to fixed input ($r=0.90, p < 0.01$ for numbers; $r=0.88, p < 0.01$ for ratios). This tight linkage suggests that mitigating the misguidance effect may be a promising direction for improving the bug-finding efficacy of LLM-generated test suites.

In real-world scenarios, the fixed version of a buggy program is unavailable for test generation. Therefore, the superior performance of tests generated from fixed code serves not as a practical baseline, but as an oracle to demonstrate the performance degradation caused by the buggy code itself. Our analysis shows that the buggy code is a primary obstacle to effective test generation, a factor that must be considered in realistic evaluations. A critical implication of these findings is that prior work \cite{OnLLMUTGenEvalASE2024, ChatVSSBSTTSE, TestGenLLMandMutIST} that evaluated LLM-based test generation using only bug-free code likely overestimated the true bug detection capabilities of these models.

\subsection{Sequence Score}

To provide model-internal evidence for the misguidance effect, we analyze the sequence score an LLM assigns to each generated test. This score reveals the model's probabilistic preference \cite{LMareFew-Shot-Learner, DetectGPT, Holtzman2020TheCuriousCase} for a given output: if a bug in the input skews the model's generation preferences, misguided tests should receive higher sequence scores when conditioned on buggy code, while effective tests should score higher when conditioned on fixed code.

To conduct this experiment, we used three SOTA open-source models as evaluators: DeepSeek-V3 \cite{deepseekai2025deepseekv3technicalreport}, Qwen3-Coder-Plus \cite{alibaba2025qwen3coder}, and GPT-OSS-120B \cite{openai2025gptoss120bgptoss20bmodel}, as sequence scores are not accessible from the API of proprietary models. To ensure an unbiased evaluation, we employed a cross-scoring methodology where each model only scored tests generated by models from different developers. This prevents a model from favoring its own generative style, which could be influenced by shared training data or a similar training process.

The results in Table~\ref{tab:sequence_score} confirm that buggy source code skews the model's generation preferences toward misguided tests that assert buggy behavior, as evidenced by two consistent trends. First, misguided tests receive higher average sequence scores when conditioned on buggy code than on fixed code; for example, with DeepSeek-V3 as evaluator, the average drops from $-1.21$ to $-1.31$ when switching from buggy to fixed input, with similar drops for Qwen3-Coder-Plus ($-1.59$ to $-1.75$) and GPT-OSS-120B ($-2.29$ to $-2.43$). Conversely, effective tests score higher when conditioned on fixed code; with DeepSeek-V3, the average rises from $-1.30$ to $-1.23$, and the same pattern holds for Qwen3-Coder-Plus ($-1.75$ to $-1.64$) and GPT-OSS-120B ($-2.50$ to $-2.42$). Together, these trends indicate that the behavioral differences observed earlier stem from the model being internally misguided by the buggy implementation.

\begin{tcolorbox}[
    colback=black!5,
    colframe=black!20,
    boxrule=1pt,
    arc=2mm,
    title=RQ1 Answer,
    halign title=center,
    fonttitle=\bfseries\small,
    boxsep=1pt,
    left=3pt,
    right=3pt,
    top=3pt,
    bottom=3pt,
    toptitle=1pt,
    bottomtitle=1pt,
    lefttitle=3pt,
    righttitle=3pt,
]
\textit{Buggy code has a severe, twofold negative impact on LLM-generated tests. It misleads models into producing misguided tests that validate the bug and simultaneously suppresses effective tests that would detect it. We further corroborate this with model-internal evidence using sequence scores, showing that the model's preference is skewed toward tests that assert the buggy behavior in the prompt. Moreover, our correlation analysis in Section~\ref{subsec:misguidance_effect_statistics} suggests that models with stronger code comprehension capabilities can be more susceptible to misguidance, and that reductions in misguided tests are correlated with increases in effective tests.}
\end{tcolorbox}



\section{RQ2: Mitigating the Misguidance Effect with a Specification-Based Method}
In RQ2, we evaluate our proposed method for mitigating the misguidance effect from buggy code.

\begin{table}[htbp!]
\centering
\tiny
\setlength{\tabcolsep}{2.5pt}
\caption{Misguided (M) and effective (E) test comparison for four inputs: only an LLM-generated docstring from buggy code, only buggy code, neither code nor docstring, and the LLM-generated docstring with buggy code. Results shown as counts (\#) and percentages (\%).}
\label{tab:docstring_improve}
\begin{tabular}{l|cc|cc|cc|cc|cc|cc|cc|cc}
\toprule
\multirow{4}{*}{\textbf{Model}} & \multicolumn{4}{c|}{\textbf{Only Doc. (Ours)}} & \multicolumn{4}{c|}{\textbf{Only Code}} & \multicolumn{4}{c|}{\textbf{No Code or Doc.}} & \multicolumn{4}{c}{\textbf{Doc. W/ Code}} \\
\cmidrule{2-17}
& \multicolumn{2}{c|}{\emph{M}} & \multicolumn{2}{c|}{\emph{E}} & \multicolumn{2}{c|}{\emph{M}} & \multicolumn{2}{c|}{\emph{E}} & \multicolumn{2}{c|}{\emph{M}} & \multicolumn{2}{c|}{\emph{E}} & \multicolumn{2}{c|}{\emph{M}} & \multicolumn{2}{c}{\emph{E}} \\
\cmidrule(lr){2-3} \cmidrule(lr){4-5} \cmidrule(lr){6-7} \cmidrule(lr){8-9} \cmidrule(lr){10-11} \cmidrule(lr){12-13} \cmidrule(lr){14-15} \cmidrule(lr){16-17}
& \textbf{\#} & \textbf{\%} & \textbf{\#} & \textbf{\%} & \textbf{\#} & \textbf{\%} & \textbf{\#} & \textbf{\%} & \textbf{\#} & \textbf{\%} & \textbf{\#} & \textbf{\%} & \textbf{\#} & \textbf{\%} & \textbf{\#} & \textbf{\%} \\
\midrule
Gemini 2.5 Pro & 110 & 2.77 & 198 & 4.98 & 173 & 5.27 & 98 & 2.98 & 49 & 1.33 & 124 & 3.37 & 165 & 4.89 & 94 & 2.78 \\
Gemini 2.5 Flash & 105 & 2.39 & 202 & 4.60 & 114 & 3.22 & 140 & 3.96 & 45 & 1.02 & 82 & 1.86 & 117 & 3.23 & 148 & 4.08 \\
Gemini 2.5 Flash (Reason) & 126 & 2.44 & 179 & 3.47 & 155 & 3.92 & 99 & 2.50 & 75 & 1.42 & 89 & 1.68 & 209 & 4.92 & 84 & 1.98 \\
Claude 4 Sonnet & 147 & 2.85 & 249 & 4.82 & 149 & 3.55 & 88 & 2.09 & 58 & 1.39 & 107 & 2.56 & 176 & 3.74 & 148 & 3.14 \\
Claude 4 Sonnet (Reason) & 165 & 2.91 & 257 & 4.53 & 187 & 4.14 & 88 & 1.95 & 70 & 1.56 & 112 & 2.49 & 196 & 3.90 & 143 & 2.84 \\
Grok-4 & 140 & 3.13 & 239 & 5.35 & 179 & 4.56 & 145 & 3.70 & 106 & 2.32 & 122 & 2.67 & 176 & 4.23 & 157 & 3.77 \\
Grok-3 & 55 & 2.02 & 119 & 4.37 & 71 & 2.79 & 93 & 3.66 & 41 & 1.60 & 93 & 3.64 & 85 & 3.25 & 89 & 3.40 \\
GPT-4.1 & 84 & 2.51 & 160 & 4.79 & 92 & 3.24 & 86 & 3.03 & 37 & 1.22 & 104 & 3.42 & 115 & 3.87 & 98 & 3.30 \\
GPT-O4-mini & 85 & 3.00 & 112 & 3.95 & 130 & 5.20 & 36 & 1.44 & 44 & 1.77 & 91 & 3.65 & 134 & 5.04 & 76 & 2.86 \\
DeepSeek-V3 & 68 & 2.02 & 155 & 4.61 & 76 & 2.53 & 113 & 3.76 & 75 & 2.50 & 77 & 2.57 & 114 & 2.33 & 149 & 3.04 \\
DeepSeek-R1 & 125 & 3.29 & 139 & 3.66 & 145 & 4.25 & 97 & 2.84 & 70 & 1.34 & 106 & 2.03 & 109 & 3.18 & 110 & 3.20 \\
Qwen3-Coder-Plus & 91 & 2.41 & 194 & 5.15 & 80 & 2.32 & 157 & 4.54 & 71 & 1.56 & 77 & 1.69 & 95 & 2.70 & 135 & 3.83 \\
Qwen3-Plus & 168 & 3.19 & 225 & 4.28 & 239 & 4.92 & 114 & 2.35 & 96 & 1.98 & 56 & 1.16 & 208 & 4.10 & 154 & 3.03 \\
\midrule
Average & 113.00 & 2.69 & 186.77 & 4.50 & 137.69 & 3.84 & 104.15 & 2.98 & 64.38 & 1.62 & 95.38 & 2.52 & 146.08 & 3.80 & 121.92 & 3.17 \\
\bottomrule
\end{tabular}
\vspace{-1.0em}
\end{table}

\begin{table}[htbp!]
\centering
\tiny
\caption{Number of focal methods with at least one misguided (M) or effective (E) test for four inputs: only an LLM-generated docstring from buggy code, only buggy code, neither code nor docstring, and the LLM-generated docstring with buggy code.}
\vspace{-0.4em}
\label{tab:docstring_improve_method_level}
\begin{tabular}{l|cc|cc|cc|cc}
\toprule
\multirow{2}{*}{\textbf{Model}} & \multicolumn{2}{c|}{\textbf{Only Doc. (Ours)}} & \multicolumn{2}{c|}{\textbf{Only Code}} & \multicolumn{2}{c|}{\textbf{No Code or Doc.}} & \multicolumn{2}{c}{\textbf{Doc. W/ Code}} \\
\cmidrule{2-9}
 & \emph{M} & \emph{E} & \emph{M} & \emph{E} & \emph{M} & \emph{E} & \emph{M} & \emph{E} \\
\midrule
Gemini 2.5 Pro & 53 & 86 & 89 & 45 & 27 & 80 & 85 & 42 \\
Gemini 2.5 Flash & 47 & 74 & 44 & 51 & 21 & 50 & 49 & 59 \\
Gemini 2.5 Flash (Reason) & 54 & 57 & 64 & 38 & 34 & 48 & 65 & 33 \\
Claude 4 Sonnet & 52 & 85 & 59 & 41 & 26 & 58 & 59 & 59 \\
Claude 4 Sonnet (Reason) & 49 & 83 & 63 & 42 & 33 & 64 & 63 & 58 \\
Grok-4 & 53 & 93 & 84 & 60 & 34 & 70 & 76 & 65 \\
Grok-3 & 32 & 60 & 46 & 43 & 26 & 59 & 44 & 43 \\
GPT-4.1 & 42 & 73 & 56 & 45 & 24 & 52 & 61 & 48 \\
GPT-O4-mini & 49 & 61 & 72 & 29 & 22 & 53 & 69 & 41 \\
DeepSeek-V3 & 39 & 65 & 42 & 51 & 28 & 57 & 40 & 60 \\
DeepSeek-R1 & 54 & 68 & 69 & 45 & 36 & 42 & 59 & 53 \\
Qwen3-Coder-Plus & 42 & 73 & 43 & 58 & 26 & 42 & 44 & 55 \\
Qwen3-Plus & 53 & 75 & 81 & 43 & 41 & 46 & 75 & 58 \\
\midrule
Average & 47.62 & 73.31 & 62.46 & 45.46 & 29.08 & 55.46 & 60.69 & 51.85 \\
\bottomrule
\end{tabular}
\vspace{-1.0em}
\end{table}

\subsection{Applying Our Approach to Buggy Code}
\label{subsec:4.1_base_prompt}

In this section, we evaluate the effectiveness of our approach by comparing it against three alternative configurations that isolate the effects of code removal and specification replacement. Tables~\ref{tab:docstring_improve} and~\ref{tab:docstring_improve_method_level} present the results for the following settings:
\begin{itemize}[leftmargin=10pt]
    \item \textbf{Our Approach (\textsc{Only Doc.}):} The prompt removes the buggy implementation and uses only the LLM-generated docstring as the behavioral input.
    \item \textbf{Direct Code-Based Approach (\textsc{Only Code}):} The prompt contains the code under test as the sole behavioral input, following prior LLM-based test-generation studies~\cite{EmpEvalforLLMUTAutoGenTSE, EvalImprChatgptUTGenFSE2024, CODAMOSAICSE2023}.
    \item \textbf{Context-Deprivation Approach (\textsc{No Code or Doc.}):} The prompt contains only the focal method contexts with no code or docstring, assessing whether the improvement comes merely from withholding implementation details rather than from replacing them with an LLM-generated docstring. This setting also reflects baselines adopted in prior test-generation studies~\cite{Doc2OracLL, TOGA, TOGLL}.
    \item \textbf{Supplementary Approach (\textsc{Doc. W/ Code}):} The prompt includes the LLM-generated docstring alongside the buggy code as additional context, following Yuan et al.'s setting~\cite{EvalImprChatgptUTGenFSE2024}.
\end{itemize}

Compared to the \textsc{Only Code} baseline, our approach significantly improves test quality: the average number of misguided tests across all models drops from 137.69 to 113.00, a reduction of 24.69 tests (-1.15\,pp). It also substantially improves bug detection, increasing the average number of effective tests from 104.15 to 186.77, an increase of 82.62 tests (+1.52\,pp). These results suggest that focusing the LLM on the specification, rather than the buggy code, mitigates misguidance and yields more useful bug-finding tests.

For the \textsc{No Code or Doc.} baseline, both misguided and effective tests are substantially reduced compared with our approach: misguided tests drop from 113.00 to 64.38 (-1.07\,pp), while effective tests drop from 186.77 to 95.38 (-1.98\,pp). Moreover, a closer examination shows only 56.69\% of its generated tests pass on either the buggy or fixed version, compared with 81.05\% for our approach. This suggests that simply removing the code under test without providing behavioral information leaves the LLM uncertain about the expected behavior, causing many tests to align with neither the buggy nor the fixed version, making this baseline unreliable for effective test generation.

The \textsc{Doc. W/ Code} approach is substantially less effective than ours: on average, it produces 33.08 more misguided tests (+1.11\,pp) and, crucially, 64.85 fewer effective tests (-1.33\,pp). Relative to the \textsc{Only Code} baseline, it offers only a marginal gain, generating 17.77 more effective tests (+0.19\,pp) while also producing 8.39 more misguided tests. This suggests that adding the LLM-generated specification as context is insufficient.

The method-level results in Table~\ref{tab:docstring_improve_method_level} further corroborate these findings, showing a similar trend at the focal-method level. Taken together, these results demonstrate that neither removing the code under test nor adding an LLM-generated docstring alone is sufficient to both mitigate the misguidance effect and improve the number of effective tests. Instead, the two components must be combined: the buggy implementation should be removed from the behavioral input and replaced with the generated specification.




\begin{table}[htbp!]
\centering
\tiny
\caption{Comparison of misguided (M) and effective (E) test results between our \textsc{Base Docstring Prompt}, \textsc{Advanced Docstring Prompt}, and \textsc{Advanced Test Prompt}, grouped by model type. Results are presented as the number of tests (\#) with the corresponding percentage (\%).}
\label{tab:advanced_docstring}
\begin{tabular}{l|cc|cc|cc|cc|cc|cc}
\toprule
\multirow{4}{*}{\textbf{Model}} & \multicolumn{4}{c|}{\textbf{Base Docstring Prompt}} & \multicolumn{4}{c|}{\textbf{Advanced Docstring Prompt}} & \multicolumn{4}{c}{\textbf{Advanced Test Prompt}} \\
\cmidrule{2-13}
& \multicolumn{2}{c|}{\emph{M}} & \multicolumn{2}{c|}{\emph{E}} & \multicolumn{2}{c|}{\emph{M}} & \multicolumn{2}{c|}{\emph{E}} & \multicolumn{2}{c|}{\emph{M}} & \multicolumn{2}{c}{\emph{E}} \\
\cmidrule(lr){2-3} \cmidrule(lr){4-5} \cmidrule(lr){6-7} \cmidrule(lr){8-9} \cmidrule(lr){10-11} \cmidrule(lr){12-13}
& \textbf{\#} & \textbf{\%} & \textbf{\#} & \textbf{\%} & \textbf{\#} & \textbf{\%} & \textbf{\#} & \textbf{\%} & \textbf{\#} & \textbf{\%} & \textbf{\#} & \textbf{\%} \\
\midrule
Gemini 2.5 Flash & 105 & 2.39 & 202 & 4.60 & 78 & 1.67 & 209 & 4.46 & 139 & 3.29 & 158 & 3.74 \\
Claude 4 Sonnet & 147 & 2.85 & 249 & 4.82 & 98 & 1.77 & 359 & 6.49 & 213 & 3.90 & 120 & 2.20 \\
Grok-3 & 55 & 2.02 & 119 & 4.37 & 52 & 1.76 & 130 & 4.39 & 88 & 2.84 & 115 & 3.72 \\
GPT-4.1 & 84 & 2.51 & 160 & 4.79 & 72 & 2.02 & 200 & 5.61 & 114 & 3.45 & 99 & 3.00 \\
DeepSeek-V3 & 68 & 2.02 & 155 & 4.61 & 52 & 1.54 & 138 & 4.09 & 147 & 2.42 & 184 & 3.03 \\
Qwen3-Coder-Plus & 91 & 2.41 & 194 & 5.15 & 89 & 2.20 & 215 & 5.31 & 177 & 3.30 & 151 & 2.82 \\
\midrule
Base Average & 91.67 & 2.37 & 179.83 & 4.72 & 73.50 & 1.83 & 208.50 & 5.06 & 146.33 & 3.20 & 137.83 & 3.09 \\
\midrule
\midrule
Gemini 2.5 Pro & 110 & 2.77 & 198 & 4.98 & 90 & 2.06 & 232 & 5.30 & 202 & 5.34 & 116 & 3.06 \\
Gemini 2.5 Flash (Reason) & 126 & 2.44 & 179 & 3.47 & 114 & 2.03 & 238 & 4.24 & 169 & 3.85 & 91 & 2.07 \\
Claude 4 Sonnet (Reason) & 165 & 2.91 & 257 & 4.53 & 108 & 1.72 & 348 & 5.55 & 256 & 4.47 & 111 & 1.94 \\
Grok-4 & 140 & 3.13 & 239 & 5.35 & 106 & 1.96 & 288 & 5.33 & 184 & 3.70 & 261 & 5.25 \\
GPT-O4-mini & 85 & 3.00 & 112 & 3.95 & 49 & 1.51 & 137 & 4.22 & 145 & 4.81 & 95 & 3.15 \\
DeepSeek-R1 & 125 & 3.29 & 139 & 3.66 & 113 & 2.62 & 191 & 4.43 & 187 & 3.87 & 153 & 3.17 \\
Qwen3-Plus & 168 & 3.19 & 225 & 4.28 & 115 & 1.99 & 308 & 5.33 & 245 & 4.80 & 137 & 2.68 \\
\midrule
Reasoning Average & 131.29 & 2.96 & 192.71 & 4.32 & 99.29 & 1.98 & 248.86 & 4.91 & 198.29 & 4.41 & 137.71 & 3.05 \\
\midrule
Total Average & 113.00 & 2.69 & 186.77 & 4.50 & 87.38 & 1.91 & 230.23 & 4.98 & 174.31 & 3.85 & 137.77 & 3.06 \\
\bottomrule
\end{tabular}
\end{table}

\begin{table}[htbp!]
\centering
\tiny
\caption{Number of focal methods with at least one misguided (M) or effective (E) test for the three prompts of Table~\ref{tab:advanced_docstring}.}
\label{tab:advanced_docstring_improve_method}
\begin{tabular}{l|cc|cc|cc}
\toprule
\multirow{2}{*}{\textbf{Model}} & \multicolumn{2}{c|}{\textbf{Base Docstring Prompt}} & \multicolumn{2}{c|}{\textbf{Advanced Docstring Prompt}} & \multicolumn{2}{c}{\textbf{Advanced Test Prompt}} \\
\cmidrule{2-7}
 & \emph{M} & \emph{E} & \emph{M} & \emph{E} & \emph{M} & \emph{E} \\
\midrule
Gemini 2.5 Flash & 47 & 74 & 38 & 75 & 53 & 51 \\
Claude 4 Sonnet & 52 & 85 & 42 & 105 & 69 & 55 \\
Grok-3 & 32 & 60 & 38 & 61 & 43 & 49 \\
GPT-4.1 & 42 & 73 & 35 & 87 & 57 & 44 \\
DeepSeek-V3 & 39 & 65 & 35 & 67 & 50 & 66 \\
Qwen3-Coder-Plus & 42 & 73 & 40 & 86 & 53 & 54 \\
\midrule
Base Average & 42.33 & 71.67 & 38.00 & 80.17 & 54.17 & 53.17 \\
\midrule
\midrule
Gemini 2.5 Pro & 53 & 86 & 43 & 100 & 90 & 50 \\
Gemini 2.5 Flash (Reason) & 54 & 57 & 44 & 78 & 65 & 32 \\
Claude 4 Sonnet (Reason) & 49 & 83 & 42 & 106 & 65 & 51 \\
Grok-4 & 53 & 93 & 47 & 113 & 69 & 91 \\
GPT-O4-mini & 49 & 61 & 35 & 82 & 84 & 41 \\
DeepSeek-R1 & 54 & 68 & 51 & 77 & 77 & 63 \\
Qwen3-Plus & 53 & 75 & 51 & 96 & 79 & 52 \\
\midrule
Reasoning Average & 52.14 & 74.71 & 44.71 & 93.14 & 75.57 & 54.29 \\
\midrule
Total Average & 47.62 & 73.31 & 41.62 & 87.15 & 65.69 & 53.77 \\
\bottomrule
\end{tabular}
\end{table}

\begin{table}[htbp!]
\centering
\tiny
\caption{Comparison of the ratio of ``False Positive'' tests asserting hallucinated behavior, between generating from the code, the base docstring approach, and the advanced analysis approach, grouped by model type.}
\label{tab:advanced_docstring_fail_on_both}
\begin{tabular}{l|c|c|c}
\toprule
\textbf{Model} & \textbf{Code Prompt} & \textbf{Base Docstring Prompt} & \textbf{Advanced Docstring Prompt} \\
\midrule
Gemini 2.5 Flash & 18.79\% & 17.69\% & 21.27\% \\
Claude 4 Sonnet & 13.84\% & 15.95\% & 16.01\% \\
Grok-3 & 17.81\% & 16.99\% & 21.82\% \\
GPT-4.1 & 17.42\% & 16.50\% & 19.37\% \\
DeepSeek-V3 & 19.37\% & 19.10\% & 19.66\% \\
Qwen3-Coder-Plus & 19.36\% & 19.08\% & 19.98\% \\
\midrule
Base Average & 17.77\% & 17.55\% & 19.68\% \\
\midrule
\midrule
Gemini 2.5 Pro & 15.43\% & 15.15\% & 17.22\% \\
Gemini 2.5 Flash (Reason) & 15.24\% & 15.27\% & 16.27\% \\
Claude 4 Sonnet (Reason) & 14.62\% & 15.59\% & 16.92\% \\
Grok-4 & 14.14\% & 12.09\% & 17.46\% \\
GPT-O4-mini & 9.24\% & 12.00\% & 15.78\% \\
DeepSeek-R1 & 18.76\% & 16.77\% & 15.91\% \\
Qwen3-Plus & 18.64\% & 18.26\% & 18.24\% \\
\midrule
Reasoning Average & 15.15\% & 15.02\% & 16.83\% \\
\midrule
Total Average & 16.36\% & 16.19\% & 18.15\% \\
\bottomrule
\end{tabular}
\end{table}

\subsection{Further Enhancing Our Approach}
\label{subsec:enhanced_approach}

While our method substantially improves over generating tests directly from buggy code, the remaining gap in misguided and effective tests relative to prompting with fixed code highlights opportunities for further improvement and motivates more sophisticated specification-construction methods that better capture the intended behavior. As introduced in Section~\ref{subsec:2.5_methodology}, we propose the \textsc{Advanced Docstring Prompt} to further leverage the code-comprehension capabilities of modern LLMs. We compare it against the two baselines discussed in Section~\ref{subsec:2.5_methodology}: (1) the \textsc{Base Docstring Prompt} in Figure~\ref{fig:docstring_prompt_templates}, corresponding to the base version of our approach evaluated in Section~\ref{subsec:4.1_base_prompt}; and (2) direct code-based test generation combined with the \textsc{Advanced Analysis Prompt} in Figure~\ref{fig:docstring_prompt_templates}, denoted as the \textsc{Advanced Test Prompt}, where the LLM is first instructed to analyze the code under test for potential bugs and then generate tests assuming all identified bugs have been fixed.

\begin{figure*}[t]
\centering
\begin{qualfigurebox}

\begin{minipage}[t]{0.495\linewidth}
\qualtitle{Fixed Code}
\begin{qualcodebox}[width=\linewidth]
\qline{1}{\qkw{public} \qtype{String} \qtype{generateToolTipFragment}(\qtype{String} toolTipText) \{}
\qline{2}{\hspace*{1.0em}\qkw{return} \qstr{" title=\textbackslash{}"} + \qmeth{ImageMapUtilities.htmlEscape}(toolTipText)}
\qline{3}{\hspace*{2.0em} + \qstr{"\textbackslash{}" alt=\textbackslash{}"\textbackslash{}""};}
\qline{4}{\}}
\end{qualcodebox}
\end{minipage}%
\hspace{0.006\linewidth}%
{\color{black!18}\vrule width 0.25pt}%
\hspace{0.006\linewidth}%
\begin{minipage}[t]{0.495\linewidth}
\qualtitle{Buggy Code}
\begin{qualcodebox}[width=\linewidth]
\qline{1}{\qkw{public} \qtype{String} \qtype{generateToolTipFragment}(\qtype{String} toolTipText) \{}
\qline{2}{\hspace*{1.0em}\qkw{return} \qstr{" title=\textbackslash{}"} + \qtype{toolTipText}}
\qline{3}{\hspace*{2.0em}+ \qstr{"\textbackslash{}" alt=\textbackslash{}"\textbackslash{}""};}
\qline{4}{\}}
\end{qualcodebox}
\end{minipage}

\qualdivider

\qualtitle{Base Docstring Prompt Output}
\begin{qualcodebox}
\qline{1}{\qdoc{/**}}
\qline{2}{\qdoc{ * Generates an HTML fragment for a tooltip attribute in an image map.}}
\qline{3}{\qdoc{ *}}
\qline{4}{\qdoc{ * The method constructs a standardized HTML title attribute (used for tooltips) along with an empty alt attribute. }}
\qline{5}{\qdoc{ * The resulting string is formatted as valid HTML markup that can be inserted into an image map element.}}
\qline{6}{\qdoc{ *}}
\qline{7}{\qdoc{ * @param toolTipText The text content to be displayed as the tooltip. This should be properly escaped for HTML inclusion.}}
\qline{8}{\qdoc{ * @return A string containing HTML attributes for tooltip functionality, formatted as ` title="[toolTipText]" alt=""` where}}
\qline{9}{\qdoc{ * [toolTipText] is the provided parameter value.}}
\qlinebadge{10}{\qdoc{ */}}{\badgebad{Generated 0 effective tests and 3 misguided tests}}
\end{qualcodebox}
\vspace{-0.45em}
\qualdivider

\qualtitle{Advanced Docstring Prompt Output}
\begin{qualcodebox}
\qline{1}{\qdoc{/**}}
\qline{2}{\qdoc{ * Generates an HTML fragment for a tooltip attribute (`title`) and an empty `alt` attribute. The method ensures the tooltip}}
\qline{3}{\qdoc{ * text is properly escaped to prevent HTML syntax errors or security vulnerabilities, e.g., XSS. If the input is `null`, an empty}}
\qline{4}{\qdoc{ * `title` attribute is returned.}}
\qline{5}{\qdoc{ *}}
\qline{6}{\qdoc{ * @param toolTipText The text to be displayed as a tooltip. If `null`, the `title` attribute will be empty.}}
\qline{7}{\qdoc{ * @return A formatted HTML fragment containing the `title` and `alt` attributes. The `title` attribute will contain the escaped}}
\qline{8}{\qdoc{ * version of `toolTipText`, and the `alt` attribute will always be empty. Example: ` title="escaped-text" alt=""`.}}
\qlinebadge{9}{\qdoc{ */}}{\badgegood{Generated 5 effective tests and 0 misguided tests}}
\end{qualcodebox}
\end{qualfigurebox}
\vspace{-0.6em}
\caption{Qualitative example comparing the \textsc{Base Docstring Prompt} and the \textsc{Advanced Docstring Prompt}.}
\Description{Qualitative example comparing the base and advanced docstring prompts on the same buggy method; the advanced docstring identifies the missing escaping step, yielding tests that detect the bug.}
\label{fig:qualitative_example}
\vspace{-1.5em}
\end{figure*}

Tables~\ref{tab:advanced_docstring} and~\ref{tab:advanced_docstring_improve_method} confirm the effectiveness of this advanced approach. For reasoning models, the advanced prompt yields substantial gains over the \textsc{Base Docstring Prompt}, reducing misguided tests by 32.00 on average (-0.98\,pp) and increasing effective tests by 56.15 (+0.59\,pp). The tailored multi-step prompt similarly improves base models, decreasing misguided tests by 18.17 (-0.54\,pp) and increasing effective tests by 28.67 (+0.34\,pp) over the \textsc{Base Docstring Prompt}. Our approach also significantly outperforms the \textsc{Advanced Test Prompt}. For reasoning models, our approach generates 99.00 fewer misguided tests (-2.43\,pp) and 111.15 more effective tests (+1.86\,pp) on average than the \textsc{Advanced Test Prompt}. For base models, our approach generates 72.83 fewer misguided tests (-1.37\,pp) and 70.67 more effective tests (+1.97\,pp) on average. These results reinforce the necessity of combining the advanced-analysis strategy with our specification-based approach.

Because our advanced approach prompts the LLM to identify potential bugs and propose fixes, it may hallucinate non-existent ``false-intent'' behavior or incorrect fixes, leading to ``False Positive'' tests that assert behavior present in neither the buggy nor the fixed version and therefore fail on both. We report the ratio of such tests in Table~\ref{tab:advanced_docstring_fail_on_both}. The base version of our approach does not increase this ratio. The advanced-analysis approach, in contrast, involves a trade-off: it slightly raises the ratio, from around 16\% to around 18\%, but in exchange yields considerable gains, reducing the number of focal methods with misguided tests by 6 on average (12.60\% reduction) and detecting 14 more bugs on average (18.88\% increase) compared with the base version.

Figure~\ref{fig:qualitative_example} presents a qualitative example illustrating how our advanced prompt elicits a more accurate specification docstring. In this example, the buggy code lacks robustness because it fails to properly escape input text. The docstring generated by our advanced prompt correctly identifies this gap and includes the necessary escaping step, which leads to the generation of tests that successfully detect the bug. In contrast, the docstring from the \textsc{Base Docstring Prompt} does not include this step, resulting in no effective tests being generated.

\begin{figure*}[ht]
    \centering
    \includegraphics[width=\linewidth]{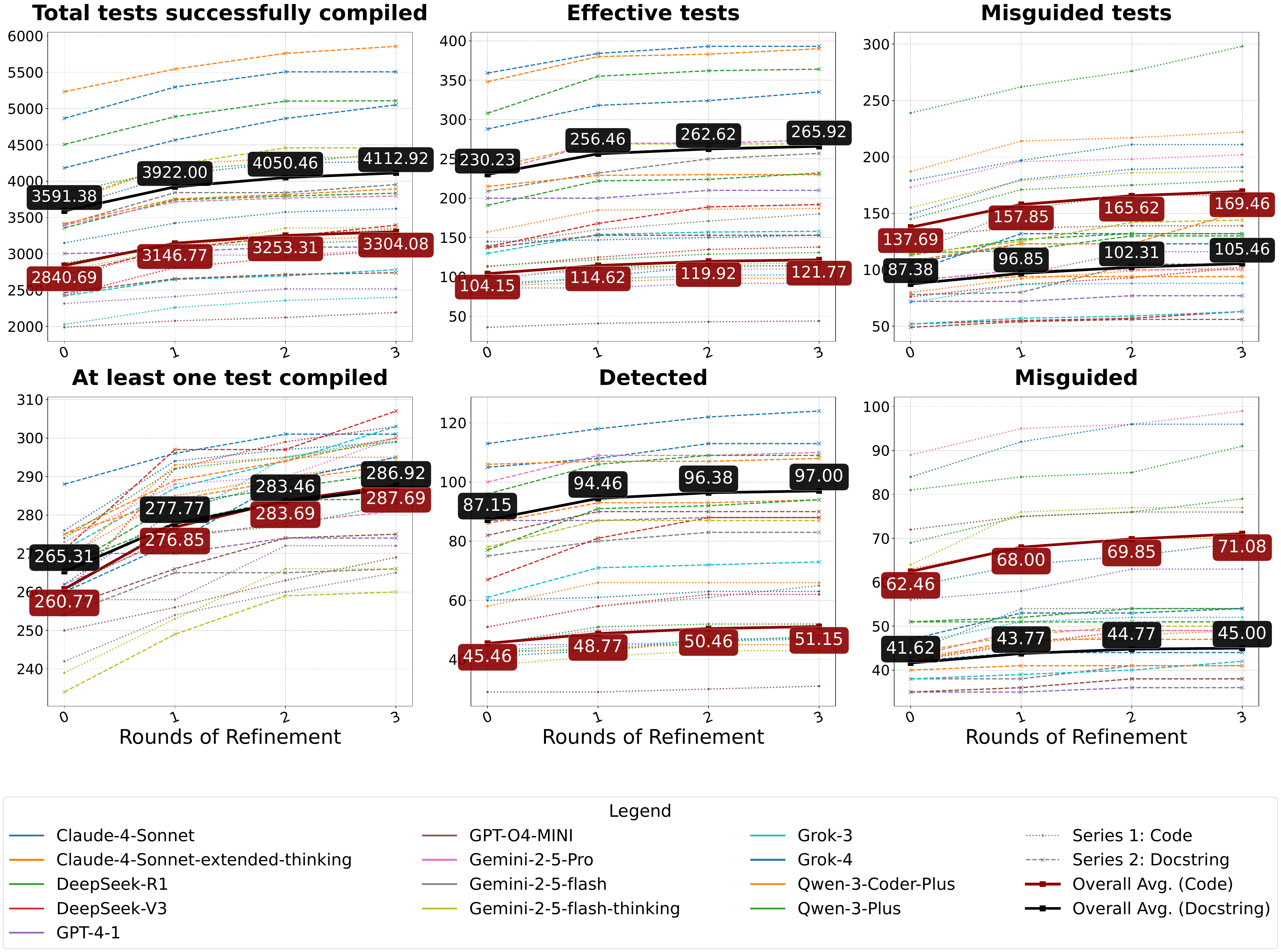}
    \vspace{-0.4em}
    \caption{Changes in the counts of compiled, effective, and misguided tests (top) and in the number of focal methods whose test suite contains at least one such test (bottom), across iterative refinement rounds, comparing buggy-code input to our docstring input.}
    \Description{Line charts showing counts of compiled, effective, and misguided tests, and of focal methods with detected bugs and misguided tests, across three refinement rounds for buggy-code and docstring inputs.}
    \label{fig:refine_comparison}
    \vspace{-1.0em}
\end{figure*}

\subsection{Impact on a Multi-Round Prompting Setup}



Recent work in unit test generation often employs multi-round prompting, in which an LLM iteratively refines tests based on feedback from sources such as execution results, human input, or other AI agents \cite{ChatUniTestFSE2024Demo, EvalImprChatgptUTGenFSE2024, testforge_multi_round}.

To investigate how the misguidance effect manifests in this multi-round, iterative setting, we adopt a feedback-driven process similar to ChatTester by Yuan et al.~\cite{EvalImprChatgptUTGenFSE2024} to evaluate our proposed approach. In this setup, if a generated test fails to execute, we extract the resulting error message and append it to the original context. This augmented prompt is then resubmitted to the LLM with instructions to revise the test suite. While specific multi-round pipelines vary in architecture, this refinement loop serves as a representative baseline for our evaluation.

Based on our results in Figure~\ref{fig:refine_comparison}, we observe that while both prompts (buggy code and specification docstring) improve compilation success at a similar rate, our specification-based approach is more robust and effective for iterative test refinement. In each round, it consistently accelerates the generation of effective tests while suppressing the creation of misguided ones.

After three rounds of refinement, the docstring input increased the average effective test count by 35.69, more than double the 17.62 increase from the buggy code input, with similar superiority at the method level (9.85 vs. 5.69). Conversely, the misguided test count grew by only 18.08 on average with the docstring input compared to 31.77 with the buggy code input, a trend that also holds at the method level (3.38 vs. 8.62). These results confirm our approach's effectiveness in a multi-round test generation setting.

\begin{table}[ht]
\vspace{-0.6em}
\centering
\caption{Cohen's $\kappa$ agreement and label-combination statistics for Gemini and Qwen.}
\vspace{-1.0em}
\label{tab:kappa_and_term_stats}

\begin{subtable}[c]{0.27\textwidth}
\centering
\scriptsize
\caption{Cohen's $\kappa$}
\vspace{-0.4em}
\label{tab:cohens_kappa_sub}
\begin{tabular}{lcc}
\hline
 & Gemini & Qwen \\
\hline
(1) & 0.82 & 0.84 \\
(2) & 0.77 & 0.79 \\
\hline
\end{tabular}
\end{subtable}
\hfill
\begin{subtable}[c]{0.72\textwidth}
\centering
\tiny
\caption{Label-combination statistics}
\vspace{-0.4em}
\label{tab:term_combination_stats_sub}
\begin{tabular}{cc|ccc|ccc}
\hline
\multirow{2}{*}{(1)}
& \multirow{2}{*}{(2)}
& \multicolumn{3}{c|}{Gemini}
& \multicolumn{3}{c}{Qwen} \\
\cline{3-8}
& & Count & Misguided & Detected & Count & Misguided & Detected \\
\hline
No  & No  & 72  & 6  & 14 & 101 & 6  & 17 \\
No  & Yes & 198 & 13 & 80 & 151 & 11 & 55 \\
Yes & No  & 48  & 24 & 6  & 66  & 23 & 14 \\
\hline
\end{tabular}
\end{subtable}
\vspace{-1.2em}
\end{table}

\subsection{How Does the Quality of the Docstrings Affect the Quality of Tests?}

To complement our empirical results, we conduct a manual inspection to assess the quality of the docstrings generated with the \textsc{Advanced Docstring Prompt} and how they affect downstream tests. As discussed in Section~\ref{subsec:2.5_methodology}, the inspection covers the two models with the largest and smallest reductions in misguided test suites under our approach: Gemini 2.5 Pro and Qwen3-Coder-Plus. Following common practice for qualitative coding~\cite{saldana2021coding, oconnor2020intercoder}, one author and one external annotator, both with substantial software-development experience, independently labeled whether each docstring (1) preserves the original bug, (2) describes the corrected behavior. We compute Cohen's $\kappa$ separately for each label and model using the initial independent annotations, and report the results in Table~\ref{tab:cohens_kappa_sub}. Disagreements were then resolved through discussion~\cite{chinh2019ways, huberman2014qualitative}, and the consensus labels were used for the final analysis. Following Landis and Koch's interpretation~\cite{landis1977measurement}, the $\kappa$ values indicate ``almost perfect agreement'' ($0.82, 0.84$) for label (1) and ``substantial agreement'' ($0.77, 0.79$) for label (2).

Table~\ref{tab:term_combination_stats_sub} presents the inspection results for all label combinations. We omit the row where both labels~(1) and~(2) are ``Yes'' because its counts are zero. This is expected: if a docstring preserves the original bug from the buggy method, it is unlikely to also describe the corrected behavior needed to fix that bug. We summarize our key findings from these results below:

\textbf{Our approach effectively reduces bug propagation into generated docstrings and recovers a substantial amount of correct behavior. }
Out of 318 generated docstrings per model, only 48 Gemini-generated docstrings (15.09\%) and 66 Qwen-generated docstrings (20.75\%) preserve the original bug. Meanwhile, 198 Gemini-generated docstrings (62.26\%) and 151 Qwen-generated docstrings (47.48\%) recover the correct behavior needed to fix the bug of the focal method.

\textbf{Docstrings without buggy behavior lead to significantly fewer misguided test suites, while docstrings that recover correct behavior contribute to substantially more detected bugs. }
For Gemini, the 270 bug-free docstrings account for only 19/43 misguided focal methods, while the 48 bug-preserving docstrings account for 24/43; the 198 docstrings that recover the correct behavior contribute 80/100 detected bugs, compared with 20/100 from the remaining 120 docstrings. Similarly for Qwen: the 252 bug-free docstrings account for 17/40 misguided focal methods versus 23/40 for the 66 bug-preserving docstrings, and the 151 docstrings that recover the correct behavior contribute 55/86 detected bugs versus 31/86 from the remaining 167 docstrings.

\begin{tcolorbox}[
    colback=black!5,
    colframe=black!20,
    boxrule=1pt,
    arc=2mm,
    title=RQ2 Answer,
    halign title=center,
    fonttitle=\bfseries\small,
    boxsep=1pt,
    left=3pt,
    right=3pt,
    top=3pt,
    bottom=3pt,
    toptitle=1pt,
    bottomtitle=1pt,
    lefttitle=3pt,
    righttitle=3pt,
]
\textit{Our specification-based method is a highly effective strategy for mitigating the misguidance effect. By replacing the buggy code with an LLM-generated specification, we significantly reduce misguided test generation while substantially increasing the effective test count. This approach outperforms prior methods that merely supplement buggy code with a specification. We also introduce an advanced, analysis-driven prompting strategy that constructs a more accurate specification, further reducing misguidance and improving bug detection efficacy. We show that our approach does not significantly increase the generation of tests that assert hallucinated behavior. In multi-round, interactive settings, our approach is more robust, accelerating bug detection while resisting error accumulation. Finally, our manual inspection shows that our approach substantially blocks bug propagation into generated docstrings and recovers the correct behavior for a large fraction of buggy focal methods. The former leads to significantly fewer misguided test suites, while the latter contributes to substantially more detected bugs. These findings show that merely removing buggy implementation details is not enough; recovering the correct behavior is essential for generating effective tests.}
\end{tcolorbox}

\section{RQ3: Effects on Bug-Free Code}

\label{subsubsec:effect_on_fixed_code}
In real-world applications, the correctness of the code under test is unknown, which means our approach must also be applicable to bug-free code. We verify this by applying our approach to the bug-free (i.e., fixed) code in our benchmark. We evaluate its applicability based on two criteria commonly adopted in prior works \cite{ChatUniTestFSE2024Demo, EvalImprChatgptUTGenFSE2024, TestGenLLMandMutIST, OnLLMUTGenEvalASE2024, EmpEvalforLLMUTAutoGenTSE}: test correctness and test coverage. The results are compared against the baseline of using bug-free code as input.

\begin{table*}[ht]
\centering
\tiny
\vspace{-1.0em}
\caption{Comparison of test correctness and coverage statistics between the baseline (Fixed Code Input) and Our Approach. Lower values are better for CFR and FAR ($\downarrow$), while higher values are better for coverage ($\uparrow$).}
\vspace{-0.8em}
\label{tab:correctness_coverage_on_fixed_code}
\begin{tabular}{l|cc|cc|cc|cc}
\toprule
\multirow{3}{*}{\textbf{Model}}
& \multicolumn{4}{c|}{\textbf{Test Correctness}}
& \multicolumn{4}{c}{\textbf{Test Coverage}} \\
\cmidrule(lr){2-5} \cmidrule(lr){6-9}
& \multicolumn{2}{c|}{\textbf{Fixed Code}}
& \multicolumn{2}{c|}{\textbf{Our Approach}}
& \multicolumn{2}{c|}{\textbf{Fixed Code}}
& \multicolumn{2}{c}{\textbf{Our Approach}} \\
\cmidrule(lr){2-3} \cmidrule(lr){4-5} \cmidrule(lr){6-7} \cmidrule(lr){8-9}
& \textbf{CFR ($\downarrow$)}
& \textbf{FAR ($\downarrow$)}
& \textbf{CFR ($\downarrow$)}
& \textbf{FAR ($\downarrow$)}
& \textbf{Line ($\uparrow$)}
& \textbf{Branch ($\uparrow$)}
& \textbf{Line ($\uparrow$)}
& \textbf{Branch ($\uparrow$)} \\
\midrule
Gemini 2.5 Pro & 16.07\% & 30.55\% & 21.78\% & 38.69\% & 79.43\% & 75.70\% & 75.39\% & 70.05\% \\
Gemini 2.5 Flash & 29.95\% & 47.59\% & 29.87\% & 51.10\% & 68.14\% & 61.11\% & 69.03\% & 62.33\% \\
Gemini 2.5 Flash (Reason) & 32.84\% & 47.80\% & 29.19\% & 48.56\% & 63.48\% & 56.51\% & 73.32\% & 68.61\% \\
Claude 4 Sonnet & 13.76\% & 29.48\% & 13.59\% & 29.69\% & 75.89\% & 72.14\% & 81.83\% & 75.99\% \\
Claude 4 Sonnet (Reason) & 19.15\% & 33.82\% & 15.43\% & 31.11\% & 73.16\% & 69.49\% & 79.84\% & 75.81\% \\
Grok-4 & 18.17\% & 31.89\% & 21.03\% & 40.57\% & 76.16\% & 71.76\% & 75.42\% & 69.63\% \\
Grok-3 & 21.85\% & 39.33\% & 18.15\% & 39.38\% & 64.67\% & 61.04\% & 70.23\% & 63.55\% \\
GPT-4.1 & 20.34\% & 36.28\% & 14.76\% & 33.78\% & 69.29\% & 64.18\% & 77.27\% & 72.31\% \\
GPT-O4-mini & 20.61\% & 32.00\% & 20.03\% & 37.26\% & 73.48\% & 68.99\% & 71.20\% & 63.74\% \\
DeepSeek-V3 & 22.04\% & 41.59\% & 20.79\% & 41.25\% & 69.43\% & 62.34\% & 73.27\% & 66.89\% \\
DeepSeek-R1 & 20.40\% & 36.17\% & 21.18\% & 38.82\% & 72.97\% & 68.31\% & 74.54\% & 68.20\% \\
Qwen3-Coder-Plus & 21.68\% & 42.71\% & 15.30\% & 35.98\% & 67.76\% & 64.12\% & 76.78\% & 73.69\% \\
Qwen3-Plus & 19.99\% & 37.65\% & 20.15\% & 41.38\% & 75.54\% & 70.28\% & 78.18\% & 73.78\% \\
\midrule
Average & 21.30\% & 37.45\% & 20.10\% & 39.04\% & 71.49\% & 66.61\% & 75.10\% & 69.58\% \\
\bottomrule
\end{tabular}
\vspace{-1.0em}
\end{table*}

First, we assess whether our approach introduces spurious issues by measuring the Compilation Failure Rate (CFR) and False-Alarm Rate (FAR). As shown in Table~\ref{tab:correctness_coverage_on_fixed_code}, our approach slightly reduces CFR by $1.20$ percentage points and slightly increases FAR by $1.59$ percentage points. Overall, for bug-free code, our approach performs comparably to prompting with the source code.

Second, we evaluate our approach's impact on test coverage, a critical metric for applications such as regression testing. The results in Table~\ref{tab:correctness_coverage_on_fixed_code} show that our specification-based approach maintains coverage comparable to the baseline of using source code directly, with minor increases of 3.61 percentage points in line coverage and 2.97 in branch coverage.

\vspace{-0.4em}
\begin{tcolorbox}[
    colback=black!5,
    colframe=black!20,
    boxrule=1pt,
    arc=2mm,
    title=RQ3 Answer,
    halign title=center,
    fonttitle=\bfseries\small,
    boxsep=1pt,
    left=3pt,
    right=3pt,
    top=3pt,
    bottom=3pt,
    toptitle=1pt,
    bottomtitle=1pt,
    lefttitle=3pt,
    righttitle=3pt,
]
\textit{Our experiments confirm that our specification-based method can be applied to both buggy and bug-free code. On bug-free code, it does not significantly increase compilation-failure or false-alarm rates compared to the code input baseline, and it maintains a comparable level of test coverage. Thus, our approach improves bug detection on faulty code without compromising the quality or reliability of tests generated for correct code, making it suitable for general use and tasks like regression testing.}
\end{tcolorbox}

\section{Threats to Validity and Limitations}

\paragraph{External Validity.}
Threats to \emph{external} validity concern the generalizability of our findings. Our study relies on the Defects4J benchmark, a common choice in prior work~\cite{UnitTestCaseGenTrans, A3TestIST, EmpEvalforLLMUTAutoGenTSE, OnLLMUTGenEvalASE2024}. We used a diverse set of 318 focal methods that span all 17 projects in Defects4J, ensuring a reasonable breadth of evaluation data. Additionally, our evaluation of 11 SOTA LLMs, representing the most advanced models at the time of our study, supports the generalizability of our results. Nevertheless, Defects4J may not fully capture two real-world scenarios. First, in proprietary or highly domain-specific systems, LLMs may lack sufficient domain knowledge or context needed to infer intended behavior, potentially producing hallucinated ``false-intent'' or incorrect bug fixes. Second, for deeply stateful or dependency-rich code, a method-level docstring may omit important class, state, or dependency information, potentially leading to unusable tests or shallower coverage. Although our prompt includes surrounding class context following Yang et al.~\cite{OnLLMUTGenEvalASE2024} and constructors of user-defined object types, this context may still be insufficient for large-scale, dependency-rich projects.

\paragraph{Internal Validity.}
Threats to \emph{internal} validity stem mainly from our experimental design and manual analysis. To mitigate experimental-design threats, we used a consistent process across all experiments: all models were evaluated with the same prompt templates in each experimental setting and identical parameter settings, and the evaluation pipeline was applied uniformly to all generated tests. We also reviewed our data collection scripts to minimize errors. For the manual inspection, annotator subjectivity may affect the labels. To mitigate this risk, we followed common qualitative-coding practice: one author and one external annotator independently labeled the docstrings, Cohen's $\kappa$ was computed on the initial labels, and disagreements were resolved through discussion until consensus was reached.

\paragraph{Construct Validity.}
Threats to \emph{construct} validity in our study stem from three main sources. First, to mitigate the inherent randomness of LLMs, we set the generation temperature to 0, where possible, to promote deterministic outputs. Second, data contamination is a potential threat, as the manually written tests in Defects4J may have been part of the models' training data. However, our proposed docstring-based approach can be seen as a form of paraphrasing, a technique commonly used to reduce the impact of data contamination~\cite{lu-etal-2024-mathgenie, yang2023rethinkingbenchmarkcontaminationlanguage, song2025llmcp}. Moreover, this prompting yielded a significant \emph{improvement} across all models rather than a drop, suggesting that the LLMs are not simply recalling memorized tests but operating on the information provided by the specifications. 
Third, our advanced-prompt approach may introduce hallucinated behavior that is then asserted by generated tests. We assess this risk by measuring ``False Positive'' tests that fail on both the buggy and fixed versions for buggy code, as well as false-alarm tests that fail on correct implementations for bug-free code. In both settings, we observe only minor increases compared with the source-code baselines, suggesting that this risk is not substantially elevated.

\section{Related Work}
\label{sec:related_work}
Traditional automatic unit test generation prior to LLMs relied on techniques like symbolic execution \cite{symbolic-execution-test-gen-1, symbolic-execution-test-gen-2}, search-based algorithms \cite{Evosuite, pynguin}, and model checking \cite{Model-checking-test-gen-2, Model-checking-test-gen-1}. However, these techniques often produce tests with low maintainability and readability compared to human-written ones, making it challenging for human developers to gain useful knowledge from them \cite{OnLLMUTGenEvalASE2024}.

Subsequently, transformer-based models \cite{vaswani2023attentionneed} demonstrated promising results in various code-related tasks, including code generation \cite{Codex}, automated program repair \cite{CoT-for-APR}, and code translation \cite{Code_translation_llm}. Early efforts to apply these models to unit test generation treated the task as a sequence-to-sequence problem—akin to machine translation—using smaller-scale architectures such as BART \cite{lewis-etal-2020-bart}, PLBART \cite{ahmad-etal-2021-PLBART}, and CodeT5 \cite{wang2021codet5} trained with code--test pairs. For instance, Tufano et al.~\cite{UnitTestCaseGenTrans} introduced a BART-based model trained specifically for unit test generation; Alagarsamy et al. proposed A3Test \cite{A3TestIST}, a PLBART-based approach that augments assertions to improve test quality; and Shin et al. \cite{codet5domainadaptUT} applied domain adaptation to CodeT5 to enhance unit test generation.

The emergence of large decoder-only LLMs, like GPT \cite{LMareFew-Shot-Learner}, capable of scaling to hundreds of billions of parameters, has enabled more natural and maintainable test generation \cite{pan2025asternaturalmultilanguageunit}. One of the earliest specialized approaches, CAT-LM \cite{CAT-LMASE2023}, trained a GPT-style model on aligned code--test data. Larger models such as GPT-4 \cite{openai2024gpt4ocard}, Claude \cite{anthropic2024claude3}, and Llama \cite{Llama}, while not explicitly trained for unit test generation, can still produce meaningful test cases when guided by well-crafted prompt instructions. Recent work has explored various prompting strategies to improve test correctness \cite{LLMforJunitTestGenEASE2024, Few-shot-test-gen, CODAMOSAICSE2023} and coverage \cite{HITS:HighCoveLLMUTGen, AutoTestImprvMetaFSE2024, coverup} with these models. Complementary studies also investigated how different fine-tuning methods further enhance LLMs' capability in unit test generation \cite{FintuningLLMforUT, storhaug2024parameterefficientfinetuninglargelanguage}.

While a significant number of prior studies have evaluated LLM-generated unit tests primarily in terms of test correctness and coverage, their core function—bug detection—has received relatively limited attention. Recent research \cite{OnLLMUTGenEvalASE2024, ChatVSSBSTTSE, TestGenLLMandMutIST} has begun to fill this gap by directly assessing and improving the bug detection capability of LLMs. However, these studies typically use correct code as input, overlooking the real-world scenarios where code under test can often be buggy. 

A few recent works have begun to evaluate unit tests generated from buggy code. Abdullin et al.~\cite{abdullin2025testwarscomparativestudy} report bug-detection rates when using buggy code as input to LLMs. Li et al.~\cite{buggy_intention_infer} propose synthesizing multiple code variants from LLM-inferred intent derived from buggy code and generating tests based on differences among these variants. They further suggest that LLMs can often infer intended behavior when buggy and fixed code differ only slightly, but the evidence is based on a small dataset (40 programs) with relatively simple tasks. Mathews et al.~\cite{buggycodeprevent} note that buggy code can prevent its own bug from being detected by the tests generated from it. Huang et al.~\cite{huang2025measuringinfluenceincorrectcode} show that tests generated from buggy code exhibit lower correctness, coverage, and bug-detection capability, and make an initial attempt to frame this phenomenon as the \emph{misguidance effect}. However, as we demonstrate in Section~\ref{subsec:metrics}, their metric does not accurately capture the existence or magnitude of misguidance. To date, no prior study has accurately quantified how buggy code misleads LLMs and proposed an effective mitigation strategy.
\section{Conclusion and Future Work}
In this paper, we provide a large-scale quantitative study of the misguidance effect in LLM-generated tests: a phenomenon where buggy code causes LLMs to misinterpret erroneous behavior as intended functionality. Using a new metric we introduce, we empirically demonstrate the effect's severe, twofold impact: it increases the generation of ``misguided tests'' that validate the bug while simultaneously suppressing ``effective tests'' that would detect it. Furthermore, we confirm this effect from a model-internal perspective by showing that buggy code skews the model's preferences toward ``misguided tests'' that assert its erroneous behavior.

To address this, we propose and validate a specification-based approach that decouples test generation from the potentially flawed implementation by using a specification constructed from the code under test as the sole behavioral input. Our experiments show that even a simple, LLM-generated docstring significantly mitigates the misguidance effect and improves bug detection. We further demonstrate that this method can be enhanced with reasoning-based code-intent analysis and serves as a more effective foundation for a multi-round, interactive prompting workflow. Finally, our approach applies to both buggy and bug-free code, achieving comparable levels of compilation failures, false-alarm tests, and test coverage on the latter.

Our work underscores the necessity of addressing the misguidance effect and demonstrates that shifting the focus from faulty implementations to intended specifications is a robust and effective path forward for LLM-based software testing. Future work includes developing more robust specification-generation approaches that reliably infer intended behavior in large-scale or domain-specific projects, potentially by combining multi-agent LLM reasoning with behavioral information from static program analysis; exploring specification formats beyond natural language, such as UML diagrams or formal specifications; and comparing LLM-generated specifications with high-quality, human-written ones to evaluate their effectiveness in mitigating misguidance and to understand the upper bound of this approach.

\begin{acks}
This work was supported by the Natural Sciences and Engineering Research Council of Canada (NSERC) under Grant RGPIN-2022-04154 and by the Connaught Fund under Grant NR-2022-23. We thank Yuliang Song for serving as an external annotator in our qualitative analysis.
\end{acks}

\section*{Data Availability}
The data used in this study is drawn from the publicly available Defects4J dataset~\citep{Defects4j}. To support reproducibility, we provide a replication package~\citep{zhao2026replication} containing our data preprocessing scripts, manual inspection results, and the source code for our evaluation pipeline and our proposed specification-based test generation approach. The package is publicly available on GitHub at \url{https://github.com/drixs2050/EvalAndMitigate} and permanently archived on Zenodo (DOI: \href{https://doi.org/10.5281/zenodo.21428153}{10.5281/zenodo.21428153}).

\bibliographystyle{ACM-Reference-Format}
\bibliography{Buggy_Code_to_Bug_Detection}










\end{document}